\documentclass[preprint, review,3p,times, 10pt]{elsarticle}

\usepackage[utf8]{inputenc}
\usepackage{multirow}
\usepackage{caption}
\usepackage{subcaption}
\usepackage{comment}
\usepackage{makecell}
\usepackage{amsmath,amssymb,amsfonts}

\usepackage[hyphens]{url}
\usepackage{hyperref}
\hypersetup{colorlinks=true,breaklinks=true}
\usepackage{setspace} %condense table text space

\usepackage{pgf-pie}
\usepackage{pgfplots}
\pgfplotsset{compat=1.8}
\usetikzlibrary{patterns}

\begin{document}
\begin{frontmatter}
% \title{Machine Learning for Chronic Wound Tissue Segmentation and Classification: Comprehensive Evaluation of Existing Techniques using A New Dataset}
\title{Deep Learning for Wound Tissue Segmentation: A Comprehensive Evaluation using A Novel Dataset}
% \author{akabircs} [orcid=0000-0002-6798-6535]

\author[1]{Muhammad Ashad Kabir\corref{cor1}}\ead{akabir@csu.edu.au}

\author[2]{Nidita Roy}\ead{u1804018@student.cuet.ac.bd}

\author[1,3]{Md. Ekramul Hossain}\ead{ekramul.hossain@sydney.edu.au}

\author[4]{Jill Featherston}\ead{Jillfeatherston@optusnet.com.au}
\author[5,6]{Sayed Ahmed}\ead{sayed@footbalancetech.com.au}

\affiliation[1]{organization={School of Computing, Mathematics and Engineering, Charles Sturt University}, city={Bathurst}, state={NSW}, postcode={2795}, country={Australia}}

\affiliation[2]{organization={Department of Computer Science and Engineering, Chittagong University of Engineering and Technology}, city={Chattogram}, postcode={4349}, country={Bangladesh}}

\affiliation[3]{organization={Complex Systems Research Group, Faculty of Engineering, The University of Sydney}, city={Darlington}, state={NSW}, postcode={2008}, country={Australia}}

\affiliation[4]{organization={School of Medicine, Cardiff University}, city={Cardiff}, state={Wales}, postcode={CF14 4YS}, country={United Kingdom}}

\affiliation[5]{organization={Principal Pedorthist, Foot Balance Technology Pty Ltd}, city={Westmead}, state={NSW}, postcode={2145}, country={Australia}}
\affiliation[6]{organization={Offloading Clinic, High-Risk Foot Services, Nepean Hospital}, city={Kingswood}, state={NSW}, postcode={2750}, country={Australia}}

\cortext[cor1]{Corresponding author: Charles Sturt University, Panorama Ave, Bathurst, NSW 2795, Australia. Ph. +61263386259, akabir@csu.edu.au}

\journal{'}
% \maketitle

\begin{abstract}
Chronic wounds represent a major global health challenge, imposing substantial economic and social burdens worldwide. Traditional methods of visually inspecting and identifying wound tissue are often time-intensive, costly, inaccurate, imprecise, and inconsistent, with a high degree of variability between clinicians. Deep learning (DL) techniques have emerged as promising solutions for medical wound tissue segmentation. However, a notable limitation in this field is the lack of publicly available labelled datasets and a standardised performance evaluation of state-of-the-art DL models on such datasets. This study addresses this gap by comprehensively evaluating various DL models for wound tissue segmentation using a novel dataset.
We have curated a dataset comprising 147 wound images exhibiting six tissue types: slough, granulation, maceration, necrosis, bone, and tendon. The dataset was meticulously labelled for semantic segmentation employing supervised machine learning techniques. Three distinct labelling formats were developed -- full image, patch, and superpixel. Our investigation encompassed a wide array of DL segmentation and classification methodologies, ranging from conventional approaches like UNet, to generative adversarial networks such as cGAN, and modified techniques like FPN+VGG16. Also, we explored DL-based classification methods (e.g., ResNet50) and machine learning-based classification leveraging DL features (e.g., AlexNet+RF).
In total, 82 wound tissue segmentation models were derived across the three labelling formats. Our analysis yielded several notable findings, including identifying optimal DL models for each labelling format based on weighted average Dice or F1 scores. Notably, FPN+VGG16 emerged as the top-performing DL model for wound tissue segmentation, achieving a dice score of 82.25\%. This study provides a valuable benchmark for evaluating wound image segmentation and classification models, offering insights to inform future research and clinical practice in wound care. The labelled dataset created in this study is available at \url{https://github.com/akabircs/WoundTissue}.
\end{abstract}


\end{frontmatter}

\section{Introduction}
Chronic wounds, including pressure ulcers, venous ulcers, diabetic foot ulcers (DFUs), infections, and other long-lasting injuries, pose a significant global health issue, carrying immense economic and social burdens~\citep{olsson2019humanistic}. These wounds often arise from conditions such as diabetes, vascular disease, immobility, and pressure injuries and tend to persist in an inflammatory state, leading to prolonged healing times, increased risk of infection, and high rates of morbidity and mortality~\citep{falanga2022chronic}. In the United States alone, over 8.2 million individuals suffer from chronic wounds, accounting for over \$28 billion in annual healthcare costs~\citep{sen2023human}. The burden is equally severe in Australia, where approximately 420,000 people are affected annually, with treatment expenses exceeding AUD 3.5 billion~\citep{australia2022pre}. Particularly concerning is the rising incidence of DFUs~\citep{lazzarini2023global}, a prevalent complication of diabetes mellitus (DM)~\citep{sun2022idf} that affects approximately 18.6 million people annually, with 34\% of diabetic patients developing a DFU in their lifetime. These ulcers carry an elevated risk of severe complications, such as lower extremity amputations (LEAs), with infection and severe cases leading to amputations in up to 20\% of instances~\citep{armstrong2023diabetic,edmonds2021current}. Consequently, chronic wounds represent a critical healthcare challenge, underscoring the need for effective management to alleviate both individual and global impacts.

One critical aspect of wound management is accurately identifying and analysing various wound tissue types~\citep{powers2016wound,bandyk2018diabetic,young2015accurate,whitehead2017identifying}. Wounds typically contain multiple tissue types, each playing a distinct role in healing. Granulation tissue, indicative of new connective tissue formation, marks a positive healing stage, while necrotic tissue, consisting of dead cells, impedes wound closure and requires debridement~\citep{alcantara2023identification}. Slough, an intermediate type containing dead cells and bacteria, also hampers healing, while macerated tissue signals excess moisture and may require specialised dressings~\citep{cutting2002maceration}. In more severe cases, wounds expose deeper structures like tendons and bones, which increases infection risks and necessitates advanced treatment strategies~\citep{clerici2010use}. Accurately identifying and quantifying wound tissue types is crucial for monitoring healing progress and ensuring comprehensive, detailed documentation, as emphasised by both the Australian guidelines for wound management~\citep{australia2016standards,chen2022australian} and the International Working Group on the Diabetic Foot (IWGDF) guidelines~\citep{schaper2024practical}. This documentation provides a chronological record of wound assessment and serves as a foundation for informed treatment decisions.

Wound tissue assessment mainly relies on manual techniques, such as visual inspection, ruler-based measurements, or digital planimetry, where clinicians subjectively estimate wound tissue composition~\citep{rogers2010digital}. This subjective evaluation can vary significantly between practitioners, making it challenging to ensure consistency and accuracy. Furthermore, chronic wounds often present irregular shapes, poorly defined boundaries, and colour heterogeneity, complicating reliable tissue assessment. The use of the red-black-yellow colour scale, while typical, lacks the precision needed for consistent and accurate wound monitoring~\citep{vermeulen2007inter}. These manual and subjective methods are time-consuming, inconsistent, and highly dependent on the clinician's experience, leading to variability in treatment approaches~\citep{wang2024wound}. Such limitations also hinder the scalability of wound care, particularly in high-demand settings. This highlights the need for automated, standardised assessment methods to enhance the efficiency, accuracy, and reliability of wound assessments and treatment planning. Such a method would enable clinicians to analyse the proportion of each tissue type quantitatively over time, facilitating more accurate assessments of healing progress and supporting timely and targeted treatment decisions~\citep{MARTIN2024}.

Recent advancements in artificial intelligence (AI), handheld imaging devices, and mobile applications hold considerable promise for transforming wound care and addressing current limitations~\citep{pappachan2022role,queen2019artificial,kabir2024mobile,rippon2024artificial}. AI-driven technologies, especially machine learning (ML) and deep learning (DL) can enhance wound assessment by automating tasks like tissue segmentation, classification, and healing prediction, thereby providing precise and reproducible analyses~\citep{lucas2021wound,chemello2022artificial,zhang2022survey,tulloch2020machine,toffaha2023leveraging,ZAHIA2020101742}. Several clinical studies have demonstrated that AI can perform wound assessment with accuracy and efficiency comparable to human specialists~\citep{jamanetworkopen2021}, often exceeding them in consistency and efficiency~\citep{reifs2023clinical,mohammed2022time}. Handheld devices, such as the Silhouette Mobile and InSight systems, offer geometric measurements of wound area and volume but cannot currently classify tissue types within the wound, limiting their utility for comprehensive wound assessment~\citep{lucas2021wound}. Meanwhile, smartphone applications like Tissue Analytics have proven effective in enhancing wound assessment and management by improving documentation, enabling remote monitoring, and reducing the burden of patient travel, but they do not support tissue classification~\citep{barakat2022reshaping}. The mobile application Swift can reduce wound assessment time by approximately half compared to traditional methods~\citep{mohammed2022time}, but it is limited to classifying only four tissue types~\citep{ramachandram2022fully} and does not support the identification of bone or tendon tissue. Despite these advancements, gaps remain in developing comprehensive wound assessment solutions that integrate the classification and segmentation of all essential tissue types into a unified framework. Many existing applications focus on a limited range of tissue types or lack access to their datasets for further research, underscoring the need for continued development to maximise AI's potential in wound care.

\begin{figure}[!ht]
    \centering
    \includegraphics[width=0.8\linewidth]{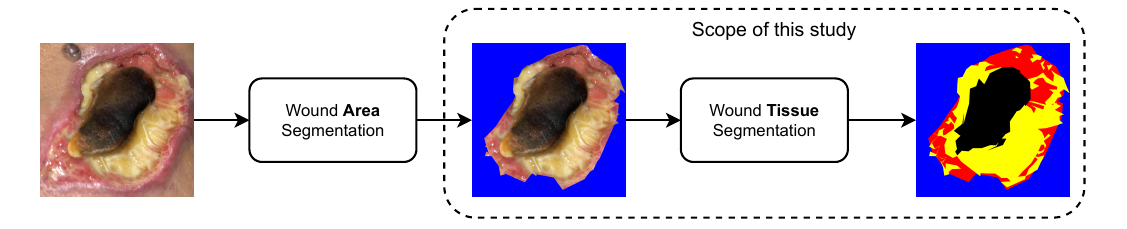}
    \caption{Overview of wound segmentation research and the scope of this study}
    \label{fig:scope}
\end{figure}
The wound segmentation problem can be broadly categorised into two key areas (illustrated in Figure~\ref{fig:scope}): wound area segmentation and wound tissue segmentation. 
\textit{Wound area segmentation} involves isolating the wound area from the background and surrounding body parts within an image. This initial segmentation step is essential as it provides a foundation for more detailed analysis, including wound tissue segmentation, wound measurement, 3D reconstruction, and wound healing assessment. A substantial body of research has been conducted in wound area segmentation, which can partly be attributed to the availability of publicly labelled datasets, such as those from the DFU segmentation challenges~\citep{YAP2024103153,kendrick2022translating} and FUSeg challenge~\citep{wang2024fuseg}. These studies have evolved from early edge-based segmentation techniques, such as active contour models~\citep{INTR5jones2000active,INTR6silveira2009comparison}, to traditional machine learning approaches using colour and texture feature-based approaches~\citep{RW3wantanajittikul2012automatic,RW4kolesnik2004segmentation,RW5kolesnik2005multi,RW6kolesnik2006robust,RW2song2012automated}, and more recently to deep learning-based approaches~\citep{RW7wang2015unified,RW10liu2017framework,INTR9chino2020segmenting,curti2022effectiveness,privalov2021software,borst2024early,RW11howard2017mobilenets,RW13wang2020fully,RW1biswas2018superpixel,zimmermann2024automated,RW39wagh2020semantic}.

In contrast, \textit{wound tissue segmentation} involves a pixel-wise classification of the various tissues (e.g., granulation, slough, etc.) found within the wound bed region. It is a more challenging task that has received less attention than wound area segmentation~\citep{ramachandram2022fully}. This disparity is primarily due to the absence of a publicly available dataset, a key barrier limiting researchers' ability to benchmark and validate new models effectively. Historically, studies in wound tissue segmentation have evolved from traditional machine learning approaches (e.g.,~\citep{RW19mukherjee2014automated,chakraborty2016telemedicine,veredas2009binary,RW23wannous2010robust}), which relied heavily on handcrafted features and simpler classifiers, to deep learning (DL) methods (e.g.,~\citep{RAJATHI2024105855,RW24niri2021superpixel,ZAHIA2020101742,RW29blanco2020superpixel}), which can learn complex representations directly from data~\citep{ZAHIA2020101742,anisuzzaman2022image,marijanovic2020systematic}. However, most DL-based studies in wound tissue segmentation have applied only a limited selection of existing DL models, primarily CNNs, without systematically comparing the latest state-of-the-art segmentation frameworks. This lack of comprehensive evaluation means that researchers have yet to determine which architectures might best address the intricate requirements of wound tissue segmentation, such as accurately segmenting various tissue types within irregular wound shapes, imbalanced datasets, and in varying lighting conditions. Furthermore, none of the studies have made their datasets publicly accessible, which creates a significant obstacle to continued research, systematic comparisons, and performance enhancement within this field.

This study systematically evaluated widely used deep learning (DL)-based segmentation and classification models for wound tissue segmentation, using a newly developed wound image dataset, named \textit{WoundTissue}. The dataset comprises 147 images annotated with six tissue types commonly found in wounds: granulation, necrosis, slough, maceration, tendon, and bone. To facilitate diverse analyses, we preprocessed the dataset into three forms: full image, patch, and superpixel. For segmentation, we categorised the methods into conventional segmentation approaches (e.g., variations of UNet~\citep{RW17ronneberger2015u}), generative adversarial networks (e.g., cGAN~\citep{MET3mirza2014conditional}), and modified segmentation architectures (e.g., UNet+VGG16, ResNet+VGG16, and FPN+VGG16). Classification was also organised into two main approaches: DL-based classifiers (e.g., VGG16, ResNet50, and InceptionV3) and machine learning (ML)-based classifiers using DL-extracted features (e.g., VGG16+RF, DenseNet201+SVM, and InceptionV3+KNN). Altogether, the varied dataset configurations led to developing and evaluating 82 unique DL models for wound tissue analyses. To our knowledge, this is the first comprehensive study to conduct a quantified comparison of 82 DL models for wound tissue segmentation and classification, offering insights into optimal model configurations for advancing wound care technology.
The contributions of this paper are summarized as follows:
\begin{itemize}

\item We introduced a new wound tissue segmentation dataset comprising six distinct types of wound tissues: granulation, necrosis, slough, maceration, tendon, and bone. Notably, this dataset includes maceration, bone, and tendon tissues previously unavailable in labelled datasets. Experts meticulously labelled these tissues, making this dataset the pioneering publicly accessible resource for labelled wound tissue imagery.

\item We comprehensively assessed state-of-the-art deep learning techniques for image segmentation using our novel wound tissue dataset. This evaluation is presented to establish a benchmark for future research in this domain, addressing a previous limitation where the analysis was restricted to only a limited number of models. 
\end{itemize}

\section{Related work} \label{sec:background-study}
This section presents an overview of recent studies that used machine learning and deep learning-based approaches specifically for wound tissue segmentation from digital images. For a comprehensive review of this topic, please refer to the references~\citep{ZAHIA2020101742,anisuzzaman2022image,marijanovic2020systematic}.

\subsection{Traditional machine learning-based approaches}
Early studies on wound tissue segmentation primarily used traditional machine learning methods that relied on handcrafted features, such as colour and texture, to segment different wound tissues. For instance, several studies~\citep{RW19mukherjee2014automated,chakraborty2016telemedicine,veredas2009binary} developed a tissue classification model using a Bayesian classifier to classify three types of tissues -- granulation, necrotic, and slough -- based on colour and texture features extracted from wound images. Similarly, \citet{RW20veredas2015wound} proposed a wound tissue classification method that applied three traditional machine learning techniques, SVM, neural network (NN), and random forest (RF), to the same three tissue types using colour and texture features. Several other studies~\citep{RW21wannous2010enhanced, RW22wannous2007supervised, RW23wannous2010robust,li2022wound} also implemented SVM-based classification methods using colour and texture features. 
In another study, \citet{RW40chang2017multimodal} introduced a multimodal sensor system for wound assessment, which used colour histograms to extract features from segmented wound images. Using the RF algorithm, these features were classified into four different wound tissues: granulation, slough, eschar, and bone/tendon. 

Despite their effectiveness to a degree, these traditional approaches face significant limitations in generalisability, primarily due to their dependence on manually designed features. Such features often fail to capture the complexity and diversity of wound tissues in real-world clinical settings. Notably, none of these studies considered all six tissue types as we have, nor did they provide publicly available datasets to facilitate reproducibility in further research.

\subsection{Deep learning-based approaches}
Recently, DL approaches have gained attention from researchers for wound tissue classification \citep{MARTIN2024}. Various approaches have been used to tackle the complexity of tissue differentiation and wound boundary delineation, each with distinct methodologies and limitations.

\subsubsection{Full image-based segmentation} 
This approach processes the entire image as a single entity. It involves applying segmentation algorithms directly to the complete image, generating a segmentation map where each pixel is classified into a specific category (e.g., wound, non-wound, or other tissue types). Full image segmentation benefits from capturing global context and relationships within the image, which can be crucial for accurate segmentation. However, it can be computationally demanding and memory-intensive, especially with high-resolution images. Popular deep learning models such as UNet~\citep{RW17ronneberger2015u} and its variants are often used for full image segmentation.

The application of \textit{full} image-based segmentation in wound analysis is demonstrated by~\citet{RW37godeiro2018chronic}, who used models like UNet, SegNet, FCN32, and FCN8 on a dataset of 30 images. This work showed promise for segmenting granulation, slough, and necrosis tissues through the limited dataset size constraints model generalisability and reliability. \citet{RAJATHI2024105855} further advanced this approach with DUTCNet, a model capable of distinguishing more diverse tissue types, including eschar and epithelial tissues, across a larger dataset of 150 images. This model demonstrated potential for capturing more nuanced tissue type differences but did not address high variability across different patient populations.
\citet{RW38sarp2021simultaneous} introduced a conditional generative adversarial network (CGAN) for simultaneous boundary detection and tissue classification. The CGAN approach uniquely facilitates more complex interactions between tissue types, although it relies on a relatively small dataset of 100 images, which could limit accuracy in more varied real-world scenarios. While full image-based segmentation provides a broad contextual overview, it requires extensive labelled data and computational resources, especially with more sophisticated models like CGANs, which are sensitive to dataset diversity and size.

\subsubsection{Superpixel-based segmentation}
In contrast to full image-based approaches, superpixel-based segmentation divides the image into clusters of pixels, referred to as superpixels, which share similar characteristics such as colour, texture, or intensity. These superpixels are used as the basic units for segmentation instead of classifying individual pixels. The main advantage of superpixel-based segmentation is the reduction in the number of elements to be classified, leading to faster and more efficient processing. Moreover, superpixels capture essential local structures and preserve boundaries more effectively than traditional pixel-wise methods. This approach effectively balances the need to capture local details and maintain computational efficiency.

\citet{RW24niri2021superpixel} applied superpixel-based segmentation with models like SegNet, UNet, and FCN variants (FCN8, FCN16, FCN32) on a dataset of 219 images, achieving precise segmentation of wound tissues such as granulation, slough, and necrosis. This approach demonstrated that superpixel-based segmentation can yield accurate results with fewer computational resources than full image processing, making it a feasible solution for clinical applications with limited hardware.
Similarly,~\citet{RW29blanco2020superpixel} leveraged superpixel-based segmentation for dermatological wound analysis, applying feature extraction models like ResNet and VGG16 on 217 images. These models enhanced classification performance by focusing on specific wound regions, although they are limited by their reliance on well-defined tissue boundaries, which are not always present in ulcer images. Superpixel-based segmentation thus balances processing efficiency and localised accuracy, but it may struggle with more heterogeneous ulcer images where tissue types blend gradually. 

\subsubsection{Patch-based segmentation}
Patch-based segmentation represents an intermediary approach wherein an image is divided into smaller patches, either overlapping or non-overlapping patches, and each patch is processed independently.
In patch-based approaches, the segmentation problem is often transformed into a classification problem by dividing the image into smaller patches and classifying each patch according to its tissue type, simplifying the segmentation task into a series of patch-level classifications.
Each patch is then classified or segmented, and the results are subsequently combined to form the final segmentation map. This approach can be advantageous in handling high-resolution images and reducing computational load as the model processes smaller sections of the image at a time. Patch-based segmentation is particularly suitable for training deep learning models on small datasets, effectively increasing the training data size without requiring additional full images.
Furthermore, it improves the focus on local features within the image. However, this approach may encounter difficulties in capturing the global context and can produce artifacts at the boundaries of the patches. Models like convolutional neural networks (CNNs)~\citep{MET7alzubaidi2021review} commonly perform segmentation tasks on patches.

One of the foundational studies using this approach was conducted by~\citet{RW25zahia2018tissue}. Wound images were partitioned into small patches of $5\times5$ pixels, and a CNN was used for tissue classification. This method allowed for high granularity in distinguishing tissue types by concentrating on local features within each patch, though the small patch size necessitates a significant number of patches per image, increasing computational demand. \citet{RW26nejati2018fine} expanded upon this approach by combining deep learning with machine learning; they divided wound images into larger $20\times20$ patches and used AlexNet to extract features, followed by an SVM classifier to categorise each patch. This hybrid model demonstrated improved accuracy in classifying tissue types, suggesting that patch size and feature extraction methods can significantly impact classification performance.

\citet{RW28rajathi2019varicose} implemented a 4-layer CNN for tissue classification of ulcer images, using the patch-based approach to improve the model's focus on subtle tissue variations within wound images. \citet{RW35garcia2018classification} further explored patch-based analysis with a 3D CNN, dividing images into $5\times5$ pixel patches to leverage depth information for pressure ulcer classification. This model emphasised spatial depth features, enhancing the classification of granulation and necrosis tissues but requiring high computational resources. Similarly, \citet{reifs2023clinical} used four commonly available CNN architectures, employing the same $5 \times 5$ patch size for their analyses.

In contrast, \citet{RW44maity2018pixel} used a pixel-based feature extraction approach, using a $9 \times 9$ mask window that scanned over each pixel in the tissue regions to extract relevant features, processed using an autoencoder-based CNN. Lastly, \citet{RW43pholberdee2018study} implemented a simpler CNN architecture that operated on larger $31 \times 31$ pixel patches. These studies highlight the diversity in tissue segmentation and classification methodologies, emphasising how variations in patch sizes and feature extraction techniques can affect model performance.

Overall, patch-based segmentation offers flexibility in wound tissue analysis by turning segmentation into a classification task, making it feasible for models to learn from limited data. Patch size, feature extraction method, and CNN architecture selection are all critical to optimising the accuracy of these models. However, this approach may sacrifice some global context, as each patch is classified independently without considering neighbouring patches, which could lead to misclassification in highly heterogeneous wound images.

\begin{table}[!htb]
\centering
    \caption{Summary of the DL-based studies}
    \label{tab:LiteratureSummary}
\resizebox{\columnwidth}{!}{
\begin{tabular}{@{\extracolsep{4pt}}cclllrc}
\hline
\multirow{2}{*}{Study} & \multicolumn{3}{c}{Method} & \multicolumn{3}{c}{Dataset}\\
\cline{2-4} \cline{5-7}
% & Approach & Input type& \multicolumn{1}{c|}{Model} & Tissue type &  \makecell[t c]{Image\\count} & \makecell[t l]{Available\\ publicly}\\ \hline\hline

% \multirow{2}{*} {Study} 
 & Approach & Input type& Model & Tissue &  \makecell[t c]{Image\\count} & \makecell[t l]{Available}\\ \hline\hline

\citep{RW37godeiro2018chronic} & Segmentation & Full image &  \makecell[t l]{UNet, SegNet, FCN32, FCN8} & G, S, N & 30 & No \\ 
% \hline
    % \cline{2-7}
\citep{RAJATHI2024105855} & Segmentation & Full image & DUTCNet & G, Es, N, Ep & 150 & No\\

% \hline 
% \cline{2-7} 
                      
\citep{RW38sarp2021simultaneous} & Segmentation & Full image & CGAN & G, S, N  & 100 & No \\ 
% \hline
\citep{RW24niri2021superpixel} & Segmentation & Superpixel & \makecell[t l]{SegNet, UNet, FCN8, FCN16, FCN32} & G, S, N &  219 & No  \\ 
  
\citep{RW29blanco2020superpixel} & Classification & Superpixel & \makecell[t l]{ResNet, VGG16, InceptionV3} & G, S, N &  217 & No \\ 
% \hline

\citep{RW25zahia2018tissue} & Classification & Patch & CNN & G, S, N   & 22 & No \\ 
                     % \hline
                    
% \multirow{6}{*}{Classification} & \citep{RW28rajathi2019varicose} & CNN  & Granulation, Slough, Necrotic, Epithelial & Patch & 1,250 & No\\ \cline{2-7}
\citep{RW28rajathi2019varicose} & Classification & Patch & CNN  & G, S, N, Ep  & 1,250 & No\\ 
% \hline
                    
\citep{RW35garcia2018classification} & Classification & Patch & 3D CNN  & G, S, N   & 193 & No \\ %RW36elmogy2018tissues
                   % \hline
                    
                 %  & \citep{RW36elmogy2018tissues} & CNN  & Granulation, Slough, Necrotic  & Patches ($7\times7$) & 3168,778 (approx.) & not available \\ \cline{2-7} 
                   
\citep{RW43pholberdee2018study} & Classification  & Patch & CNN  & G, S, N  & 180 & No \\ 
                  % \hline
                   
\citep{RW44maity2018pixel} & Classification & Pixel  & CNN & G, S, N & 250 & No \\ 
% \hline 
\citep{RW26nejati2018fine} & Classification & Patch & AlexNet+SVM & G*, S, N, Ep, I  & 350 &  No\\ 

\citep{reifs2023clinical} & Classification & Patch & \makecell[t l]{VGG16, InceptionResNetV2,\\InceptionV3, ResNet50} & G, S, N & 727 & No\\

\hline
                   
% This study & Segmentation \&\\Classification & Full image,\\Patch, \\Superpixel &  State-of-the-art\\ DL models & Granulation, Slough, Necrotic, Maceration, Tendon, Bone   & 147 & Yes\\  \hline 
                
This study & \makecell[t l]{Segmentation and\\Classification} & \makecell[t l]{Full image,\\Superpixel,\\Patch} & \makecell[t l]{State-of-the-art 82\\ DL approaches} & G, S, N,\textbf{ M}, \textbf{T}, \textbf{B}   & 147 & Yes\\  
\hline 

\hline
\multicolumn{7}{l}{G - Granulation, S - Slough, N - Necrosis, Es - Eschar, Ep - Epethelial, M - Maceration, T - Tendon, B - Bone, I - Infected}\\
\multicolumn{7}{l}{*Classified Granulation (G) tissue as Healthy Granulation, Unhealthy Granulation, and Hyper Granulation}
\end{tabular}
}
\end{table}

\subsubsection{Summary}
Table \ref{tab:LiteratureSummary} summarises the above-discussed studies related to deep learning (DL) approaches for wound tissue segmentation and classification, as the main objective of this study is to compare different DL approaches for wound tissue analysis. Notably, existing studies have focused exclusively on either segmentation or classification methods for segmenting wound tissues without considering both approaches. In contrast, our study presents a unique and comprehensive framework incorporating 82 distinct DL models for classification and segmentation approaches. 

It is also observed that most previous studies considered three tissue types: granulation, necrosis, and slough, such as \citep{RW37godeiro2018chronic,RW38sarp2021simultaneous,RW25zahia2018tissue,RW29blanco2020superpixel}. A few studies included epithelial tissue alongside these three \citep{RW26nejati2018fine,RW28rajathi2019varicose}. In contrast, our study expanded the segmentation scope to include six tissue types: granulation, necrosis, slough, maceration, tendon, and bone, excluding epithelial tissue, which is often similar to granulation tissue in terms of its role in wound healing. Moreover, none of the studies considered both bone and tendon tissues individually in their analysis, like this study.
Furthermore, none of the existing studies made their datasets publicly available, whereas we provide access to our dataset to facilitate further research in this area. 

Regarding dataset forms, most existing studies used a patch-based approach, while a few studies used full images and superpixels as inputs to their DL models. Our research incorporated all three forms of the dataset -- full images, patches, and superpixels. This comprehensive approach allows for a more robust comparison of DL models across varying input types, addressing the limitations observed in previous studies.

\section{Material and methods} \label{sec:methodology} 
\subsection{Dataset curation}
\begin{figure}[htb]
    \centering
    \includegraphics[width=1\textwidth]{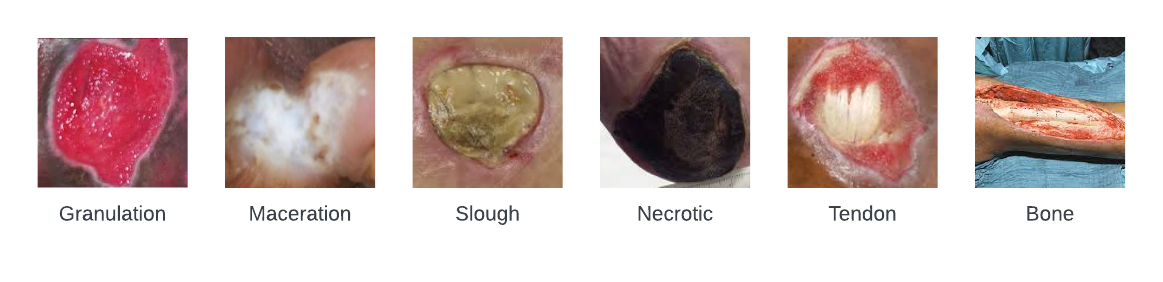}
    \caption{Different types of wound tissues in the dataset.}
    \label{fig:WoundTissues}
\end{figure}

This study created a new dataset of wound images specifically for ulcer tissue analysis. The images were sourced from multiple databases, including the DFU database \citep{MET17alzubaidi2020dfu_qutnet} and the Medetec Wound Database \citep{MET16thomas2017medetec}. However, we encountered challenges related to suboptimal image quality, which made it difficult to distinguish between various tissue types and the need to address dataset imbalance. The abundance of granulating tissue and necrotic-like erythema tissue further complicated the analysis. To maintain the correctness and reliability of the dataset, we opted to remove certain images, resulting in a final collection of 147 wound images. This dataset encompasses various types of ulcers, including pressure ulcers, diabetic foot ulcers, and venous ulcers. The wound images feature a range of tissue types, such as granulation, necrosis, slough, maceration, tendon, and bone. Figure~\ref{fig:WoundTissues} illustrates wound image samples for each of the six tissue types.

\subsection{Dataset preparation and labelling}
After collecting the raw dataset, we implemented several steps to prepare it for use in deep learning (DL) models. The first step involved segmenting the wound area, distinguishing it from healthy skin, and removing background elements. Eliminating background objects is essential for emphasising tissue features within the wound and simplifying the tasks of tissue segmentation. In this study, we manually segmented the wound areas and replaced the background with a uniform blue. This process resulted in a dataset of segmented wound images optimised for subsequent analysis.
In the second step, we applied padding to the images to ensure consistent input dimensions required by various models during training. This preprocessing step prevents unnecessary distortion, such as shrinking or stretching, and allows the images to be resized effectively before input into the models.

In the third step, we conducted detailed tissue annotation of the ulcer images, specifically delineating the various tissue types within the ulcer region. The six selected tissue classes were labelled using the following colour scheme: granulation tissue is represented in red, slough in yellow, maceration in white, necrosis in black, tendon in sky blue, and bone in cream. The background is denoted by blue. This labelling process was critical for the subsequent segmentation task, allowing us to generate precise ground truth data for model training. A team of trained annotators carefully performed the labelling process. To ensure these labels' accuracy and clinical relevance, they were thoroughly reviewed and validated by a qualified medical expert with specialised knowledge in wound care. This critical validation step was implemented to minimise labelling errors and maintain the integrity of the dataset used for developing the segmentation model. The resulting dataset is referred to as the `full image' dataset.

\begin{figure}[!htb]
     \centering
     \begin{subfigure}[b]{1\textwidth}
         \centering
         \includegraphics[width=.65\textwidth]{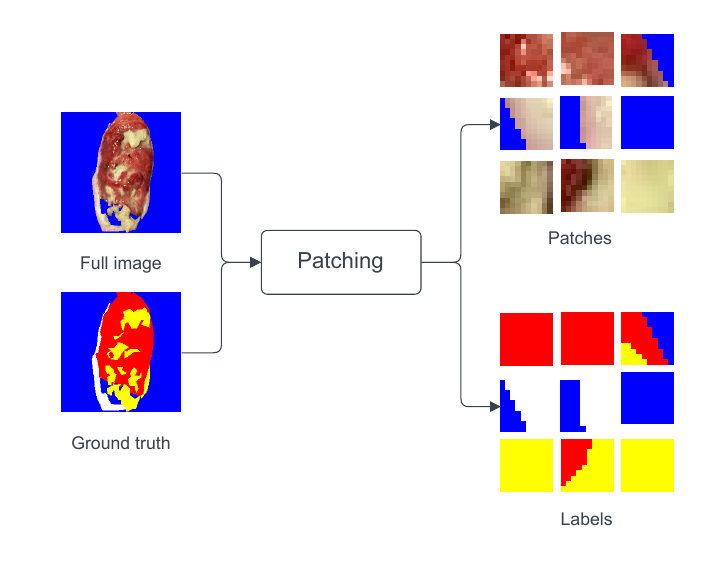}
         \caption{Patching}
         \label{fig:PatchingProcess}
     \end{subfigure}
     \hfill
     \begin{subfigure}[b]{1\textwidth}
         \centering
       \includegraphics[width=.95\textwidth]{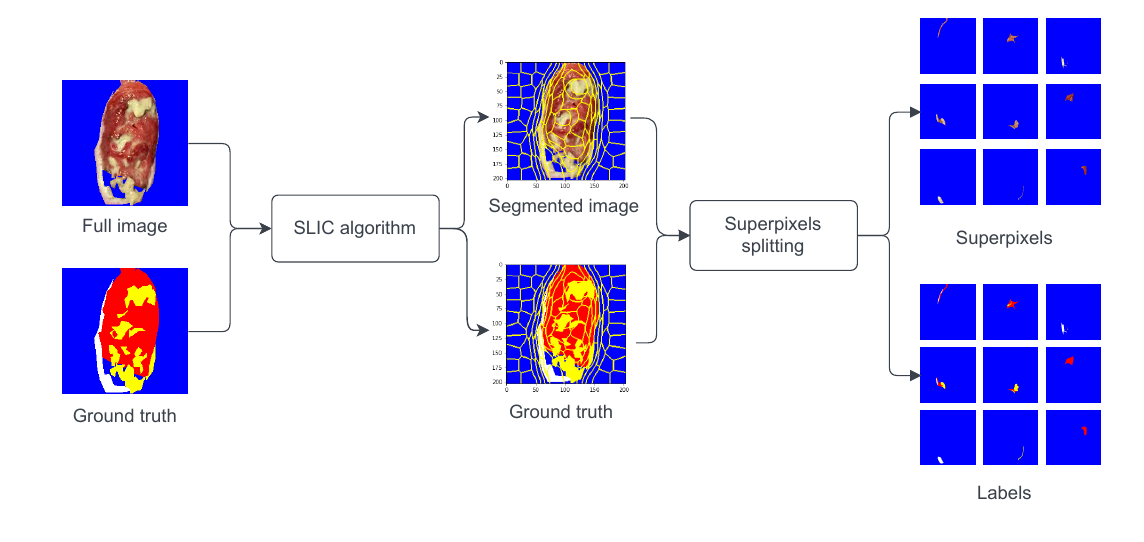}
         \caption{Superpixels}
         \label{fig:SuperpixelsProcess}
     \end{subfigure}
        \caption{Patching and superpixels}
        \label{fig:PatchSuperpixel}
\end{figure}

In the next step, we generated a patched and superpixel form of the full image dataset. The patching form of the dataset includes patches of wound images, where each wound image is divided into patches of size $10\times10$ pixels. We used the Python library \textit{patchify} \cite{online1} to split images into these smaller patches. The process for creating the patches is illustrated in Figure \ref{fig:PatchingProcess}. In contrast, the superpixel form of the dataset consists of superpixels, which are defined as groups of pixels with similar characteristics. This study used the simple linear iterative clustering (SLIC) algorithm \cite{METachanta2010slic} to generate superpixels from the segmented wound images. Each image was divided into 100 superpixels, each with an approximate size of $128 \times 128$ pixels. The process involved in creating the superpixels is shown in Figure~\ref{fig:SuperpixelsProcess}. Each patch or superpixel of an image was labelled with one of the tissue classes by comparing it to the ground truth of the corresponding original image. Patches or superpixels containing a significant presence of two or more dominant tissue labels were excluded from the dataset.

 The number of images for each tissue class across these three dataset forms -- full image (D1), patch (D2), and superpixel (D3) -- is reported in Table~\ref{tab:Datasummary}. D1 is suitable for applying segmentation techniques, D2 is appropriate for classification techniques, while D3 can be used for both segmentation and classification.

 \begin{table}[!htb]
\centering
    \caption{Summary of \textit{WoundTissue} dataset.}
    \label{tab:Datasummary}
\resizebox{1\textwidth}{!}{
\begin{tabular}{llrrrrrrr}
\hline
Dataset form & Count & Granulation & Necrosis & Slough & Maceration & Tendon & Bone & Total \\ \hline\hline

\multirow{2}{*}{Full image (D1)} & Image & 102 & 19 & 74 & 25 & 15 & 12 & 147 \\ 
& Pixel & 4,815,302 & 812,556 & 1,735,373 & 477,249 & 73,498 & 41,074 & 7,955,052\\

                        %   & Original augmented & 925 & 696 & 688 & 191 & 504 & 857 & 1382 \\ \cline{2-10} 

Patch (D2) & Image & 21,846 & 7,861 & 15,548 & 10,843 & 448 & 240 & 56,786 \\ % \cline{2-10} 
                          
Superpixel (D3) & Image & 1,802 & 420 & 1,041 & 678 & 80 & 18 & 4,039 \\ 
                           
                        %   & Superpixel augmented & 1802 & 1386 & 1463 & 1412 & 1385 & 1321 & 8769 \\ \cline{2-10} 

\hline

\hline
\end{tabular}
}
\end{table}

Each form of the dataset is split into training, validation, and testing sets, ensuring that each set across all dataset forms includes the same segmented wound images. The number of images, patches, and superpixels allocated for training, validation, and testing is detailed in Table~\ref{tab:Datasplitratio}.

\begin{table}[!htb]
\centering
    \caption{Summary of the dataset split ratio.}
    \label{tab:Datasplitratio}
\begin{tabular}{crrrr}
\hline
Dataset form & Training & Validation  & Testing & Total   \\ 
\hline\hline

D1  & 118 & 14 & 15 & 147 \\

% C2  & 1212 & 155 & 15 & 1382\\

D2  & 35196 & 9966 & 31251 & 76413\\

D3  & 2873 & 481 & 685 & 4039\\

% C5  & 7146 & 938 & 685 & 8769\\

\hline

\hline
\end{tabular}
\end{table}

\subsection{Wound tissue segmentation and classification approaches}
\begin{figure}[htb]
    \centering
    \includegraphics[width=1\textwidth]{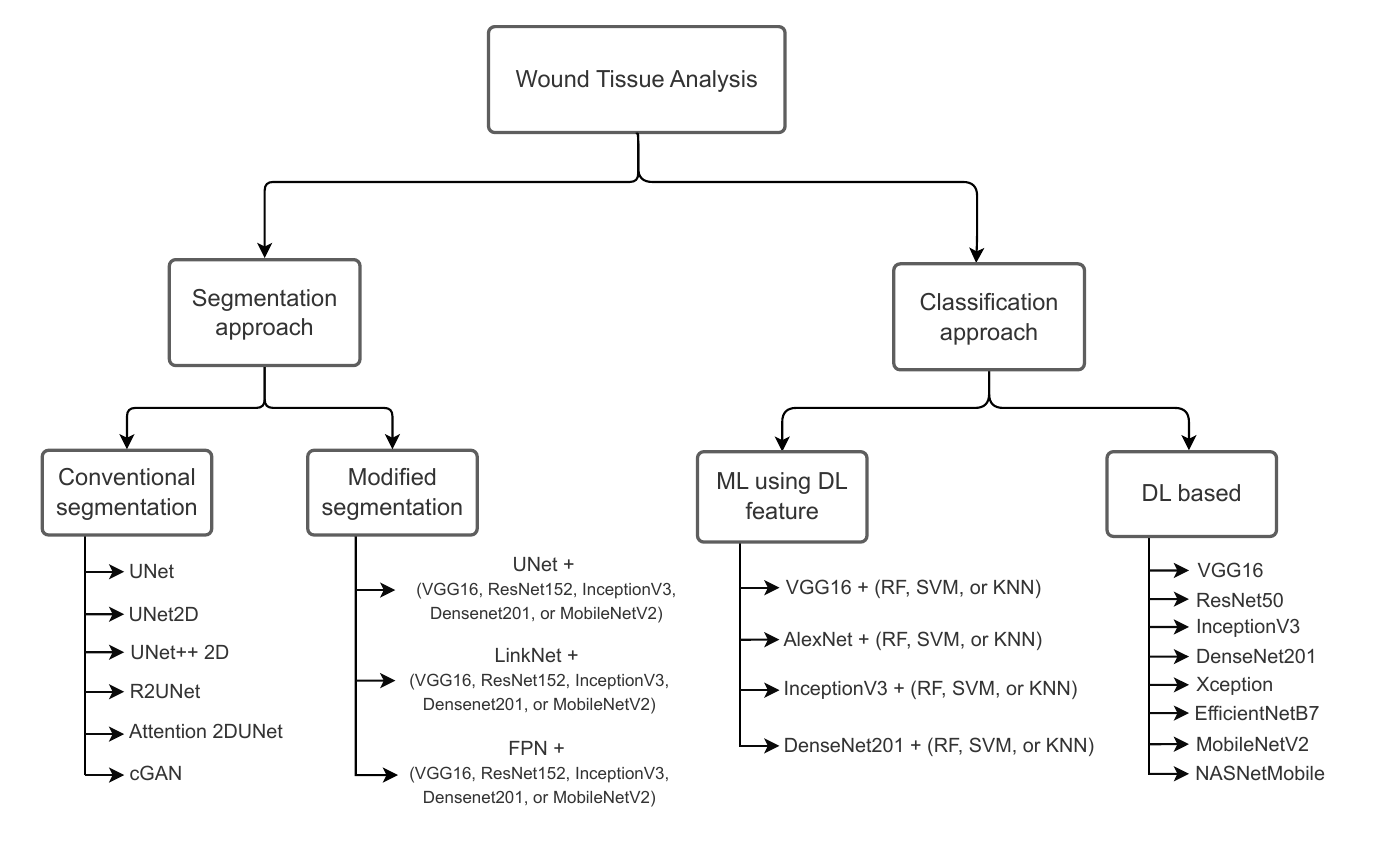}
    \caption{A taxonomy of different deep learning approaches for wound tissue classification and segmentation.}
    \label{fig:Class_Appr}
\end{figure}

This subsection describes the state-of-the-art deep learning (DL) approaches used for wound tissue segmentation and classification. A taxonomy of these approaches is illustrated in Figure~\ref{fig:Class_Appr}.   

\subsubsection{Classification approach}
In recent years, deep learning (DL) techniques have been extensively used for wound tissue classification~\citep{RW29blanco2020superpixel,RW28rajathi2019varicose,RW36elmogy2018tissues}. This study uses several prominent DL approaches for wound tissue classification, described below.

CNNs~\citep{MET7alzubaidi2021review} are among the most popular and widely used techniques in DL. The CNN architecture primarily consists of three layers: convolutional, pooling, and fully connected layers. Convolutional layers contain convolutional filters (or kernels) and generate feature maps. Each filter has weighted inputs convolved with the input volume to produce a corresponding feature map. The pooling layer reduces the dimensions of larger feature maps, resulting in smaller feature maps. Fully connected layers are located at the end of the CNN architecture and function as the classifier. Consequently, the output of these fully connected layers represents the final output of the model. 

AlexNet~\cite{MET8krizhevsky2012imagenet} is the first CNN that significantly advanced the understanding of visual data. The architecture comprises eight layers, with the first five consisting of convolutional layers followed by max-pooling layers, and the last three are fully connected layers. It uses RELU activation in each layer, except for the output layer. 
VGG~\citep{RW12simonyan2014very} is another influential CNN architecture that comprises 13 convolutional layers and three fully connected layers. It classifies images into 1,000 categories and achieved a top-5 test accuracy of 92.7\% on the ImageNet Dataset. It has several variants, including VGG11, VGG13, VGG16, and VGG19. 

ResNet~\citep{MET5he2016deep} is an advanced neural network that uses a technique known as skip connections, which allows the network to bypass one or more layers and connect directly to the output layer. This design effectively addresses the challenges of vanishing and exploding gradients. By using skip connections, the network can learn residual mappings rather than relying solely on the underlining mappings. 

Inception-V3 \cite{MET6szegedy2016rethinking} is a CNN model from the Inception family that incorporates several modifications, including the factorisation of convolutions into smaller components ($7\times7$ convolutions), label smoothing, and the use of an auxiliary classifier to propagate class information. 
DenseNet~\citep{MET9huang2017densely} is another type of CNN that uses connections between each layer and every other layer in a feed-forward manner. Xception~\citep{MET10chollet2017xception} is a variation of the Inception model that uses depth-wise separable convolutions followed by point-wise convolutions. 

MobileNetV2~\citep{MET11sandler2018mobilenetv2} is a CNN architecture that uses depth-wise separable convolutions to construct a lightweight model suitable for training on mobile devices with limited computational resources. EfficientNet~\citep{MET12tan2019efficientnet} is an enhanced version of MobileNetV2 that introduces a compound scaling module, which effectively scales depth, width, and resolution uniformly. NASNetMobile \citep{MET13tsang2020review} is another lightweight CNN architecture designed for mobile devices. It features two main components: the normal cell and the reduction cell. The normal cell generates a feature map of the same dimensions, whereas the reduction cell produces a reduced feature map, decreasing the height and width. 

In addition to the aforementioned DL-based techniques, this study uses several modified classification approaches that combine traditional machine learning (ML) algorithms with DL models. In this hybrid approach, DL models are used to extract features from the wound images, which feed into traditional ML algorithms for the classification of wound tissues. The modified approach integrates various DL models, including VGG, AlexNet, Inception, and DenseNet, along with several traditional ML models, including RF, SVM, and KNN.

\subsubsection{Segmentation approach}
This study uses several prominent DL-based techniques that have been successfully applied for segmentation in wound tissue~\citep{RW24niri2021superpixel,RW37godeiro2018chronic,RW38sarp2021simultaneous} or other domains, which are described below.

UNet~\citep{RW17ronneberger2015u} is the most widely used semantic segmentation model in biomedical image processing. It consists of an encoder that scales the features down to a smaller dimension bottleneck and a decoder that scales them back up to their original dimensions. In this study, we employed different modified versions of UNet, including UNet2D, UNet++ 2D, R2UNet, and Attention 2DUNet.

LinkNet \citep{MET14chaurasia2017linknet} is similar to UNet, with the main distinction being that LinkNet uses residual blocks in its encoder and decoder networks instead of conventional convolutional blocks. Similarly, Feature Pyramid Network (FPN)~\citep{MET15lin2017feature} also resembles UNet, but it incorporates a 1$\times$1 convolution layer and adds features rather than copying and appending them, as is done in the UNet architecture. 

Conditional generative adversarial network (cGAN) \citep{MET3mirza2014conditional} is a type of generative adversarial network (GAN). In a GAN, a generator learns to create new images, while a discriminator learns to differentiate between synthetic and real images. In cGAN, conditions are imposed on both the generator and the discriminator using information from other modalities. This enables the model to learn multimodal mappings from inputs to outputs based on diverse information sources.

In addition to the aforementioned conventional segmentation techniques, this study uses several modified segmentation techniques. In this approach, we use a combination of segmentation methods and DL models, where DL models function as encoders and conventional segmentation methods serve as base models. The modified segmentation approach incorporates the use of UNet, LinkNet, and FPN as base models, along with VGG16, ResNet152, InceptionV3, DenseNet201, and MobileNetV2 as encoders.

\subsection{Evaluation metrics}
The performance of the wound tissue segmentation and classification models is evaluated using several key metrics, including precision, recall, specificity, dice coefficient (Dice), area under the receiver operating characteristic (ROC) curve (AUC), intersection over union (IoU), and Matthews' correlation coefficient (MCC) \cite{RW25zahia2018tissue}. These metrics provide a comprehensive understanding of the models' performance, particularly in distinguishing between different tissue types.
We also computed the weighted average for each of these metrics to account for the class imbalance present in the dataset, particularly for underrepresented tissues like bone and tendon. The weighted average was calculated by multiplying the metric score for each class by the proportion of instances in that class and then summing these values. This approach ensures that the overall performance metrics are more representative of the actual class distributions in the dataset, giving a clearer picture of a model's ability to handle both dominant and rare tissue types.
The ROC curve is characterised by two parameters: the true positive rate (TPR) and the false positive rate (FPR). The other metrics are calculated based on true positive (TP), true negative (TN), false positive (FP), and false negative (FN). Definitions and equations for these metrics and detailed descriptions of how they are calculated are provided in Table~\ref{tab:Metrics}. 
\begin{table}[!htp]
\centering
    \caption{Definition and equation of the metrics.}
    \label{tab:Metrics}
\begin{tabular}{l l p{9.2cm}}
\hline
Metrics & Equation & Definition \\   
\hline\hline

Precision & $\frac{TP}{TP+FP}$ & Quantifies the accuracy of correctly identified pixels for a specific class within the segmented image.  A higher precision indicates fewer false positives, reflecting a model’s ability to precisely segment only the intended class pixels without mis-classifying other regions as part of that class.\\

\hline

\makecell[t l]{Recall/True \\Positive Rate} & $\frac{TP}{TP+FN}$ & Measures the proportion of pixels correctly identified for a particular class relative to the total number of pixels that belong to that class in the ground truth. Higher recall indicates that the model has successfully segmented a greater proportion of the actual area belonging to the target class. \\

\hline

Specificity  & $\frac{TN}{TN+FP}$ & Represents the accurately predicted negatives ratio. High specificity implies that the model effectively avoids incorrectly segmenting irrelevant areas as part of the target class.\\

\hline

Dice/F1  & $\frac{2 \times TP}{2 \times TP+FP+FN}$ & Compares the predicted segmentation and the ground truth overlap. It is defined as twice the area of true positives (TP) divided by the sum of the total number of pixels in both the predicted segmentation and the ground truth. Dice is frequently used in medical image analysis, where a slightly higher sensitivity to overlap is beneficial, especially for small and complex structures. It is also called the F1, as it balances precision and recall in a single measurement.\\

% \hline

% \makecell[t l]{False Positive\\ rate}  & $\frac{FP}{FP+TN}$ & It is the proportion of true negatives that are wrongly classified as positives.\\

\hline

IoU  & $\frac{TP}{TP+FP+FN}$ & Like Dice, IoU measures the overlap between predicted and ground truth segmentations, but it calculates this similarity differently. IoU is calculated by dividing the area of overlap (intersection) between the predicted and actual segmentation (i.e., TP) by the area covered by both (union) of these regions. It is often considered a stricter metric, as it directly measures the proportion of overlap relative to the union, making it less forgiving of false positives and false negatives.\\

\hline

MCC  & $\frac{TP\times TN-FP\times FN}{\sqrt{(TP+FP)(TP+FN)(TN+FP)(TN+FN)}}$ & Widely used to assess classification performance, particularly effective for imbalanced datasets. A high MCC score indicates strong predictive accuracy across all four components of the confusion matrix (i.e., TP, TN, FP, and FN), suggesting that the model identifies both classes without bias. MCC values range from -1 to 1, where 1 represents a perfect prediction, 0 indicates no better than random chance, and -1 corresponds to total disagreement between prediction and ground truth.\\

\hline

AUC  & -- & Measures the performance of a classification model by evaluating the area under the ROC curve. It represents the probability that a randomly selected positive instance is correctly ranked higher than a randomly selected negative instance. An AUC value closer to 1 indicates excellent model performance, while a value closer to 0.5 suggests that the model is no better than random guessing. AUC is particularly useful for assessing classifier performance in imbalanced datasets, as it considers both the true positive and false positive rates across all classification thresholds. \\

\hline

\hline
\end{tabular}
\end{table}

%MCC = (TP*TN – FP*FN) / √(TP+FP)(TP+FN)(TN+FP)(TN+FN)
\subsection{Model training}
In this study, we trained 82 distinct deep learning (DL)-based classification and segmentation models for the experimental evaluations of wound tissue segmentation. Transfer learning was used, utilising pre-trained \textit{ImageNet} weights. The models were implemented in Python using the Keras \citep{RT1chollet2015keras} and TensorFlow \citep{RT2abadi2016tensorflow} frameworks, and they were trained on a Lambda server equipped with a single NVIDIA GeForce RTX 2080 Ti GPU to enhance computational efficiency. 

Categorical cross-entropy was used as the loss function for the multi-class segmentation and classification models. It measures the difference between the predicted probability distribution of pixel classes and the true class labels (one-hot encoded) for each pixel, which can be defined as follows:
\begin{equation}
L_{\text{CCE}} = - \frac{1}{N} \sum_{i=1}^{N} \sum_{c=1}^{C} w_c \cdot y^c_i \cdot \log(\hat{y}^c_i)
\end{equation}
where \(y_i\) is the ground truth label for pixel \(i\) and class \(c\), \(\hat{y}_i\) is the predicted probability that pixel \(i\) belongs to class \(c\), \(N\) is the total number of pixels, \(C\) is the total number of classes, and \(w_c\) is the weight assigned to class \(c\), which adjusts the contribution of each class to the loss. In semantic segmentation, class weights are often used to address the class imbalance, especially when some classes have significantly more pixels than others in the dataset, like this study. The goal of calculating class weights is to give more importance to underrepresented classes during training. We assigned class weight using the equation as follows:
\begin{equation}
    % w_c = \frac{n_c}{\sum_{j=1}^{C} n_j}
    w_c = \frac{N}{C \cdot n_c}
\end{equation}
where \(n_c\) is the number of pixels for class \(c\). 

We used the \textit{Adam} optimisation algorithm to update the network parameters, which is well-regarded in stochastic optimisation for its faster convergence compared to other optimisation algorithms \citep{RT3kingma2014adam}. We trained models with an initial learning rate of \(1 \times 10^{-4}\), which was reduced by a factor of 0.1 if the validation loss did not improve for five consecutive epochs.
The total number of epochs was set to 200 with a batch size of eight; however, early stopping was implemented to terminate the training process when no improvement in validation loss was observed for more than fifteen epochs. 

We used the Intersection over Union (IoU), also called the Jaccard index, as the evaluation metric and implemented a ModelCheckpoint to save the model with the highest IoU, ensuring that the best-performing model was retained. The IoU expression presented in Table~\ref{tab:Metrics} can be adapted to calculate the weighted IoU as follows:

\begin{equation}
\text{Weighted IoU} = \frac{\sum_{c=1}^{C} n_c \cdot \frac{|y_c \cap \hat{y}_c|}{|y_c \cup \hat{y}_c|}}{\sum_{c=1}^{C} n_c} 
% \frac{}{\sum_{i=1}^{C} w_i}
\end{equation}
where,  \(y_c\) and \(\hat{y}_c\) represent the ground truth and predicted binary masks for class \(c\), respectively. The numerator computes the weighted sum of IoUs for each class. The denominator ensures normalisation by the sum of weights. $C$ is the total number of tissue classes. 

\section{Experimental evaluations} \label{sec:experiment}
The experimental evaluations in this study aimed to achieve four main objectives: (i) to provide baseline evaluation results for four segmentation and classification techniques applied to our wound dataset; (ii) to assess the impact of class weighting on the performance of classification models dealing with imbalanced data; (iii) tissue-wise performance of the selected segmentation and classification models; and (iv) to present leave-one-out cross-validation results for the selected best-performing models, thereby offering a robust estimate of their performance.

\subsection{Performance of conventional segmentation models}
This study employed six DL-based conventional segmentation models -- UNet, UNet2D, UNet++ 2D, R2Unet, Attention 2DUNet, and cGAN -- to evaluate their performance in wound tissue segmentation. Each model was applied to the full image (D1) and superpixel (D3) forms of the dataset, and the average weight of all tissue classes for the evaluation metrics was computed based on the test dataset. The results are reported in Table \ref{tab:ConvSeg}, with the best results for each dataset highlighted in bold.

The analysis revealed that the weighted average Dice score and IoU for UNet2D were the highest for D1, achieving 68.87\% and 54.22\%, respectively. Conversely, UNet++ 2D attained the highest Dice score and IoU of 68.07\% and 54.44\%, respectively, for D2. This study used the Dice score as the criterion for selecting the best segmentation model. Therefore, based on overall performance, it is evident that the UNet++ 2D model achieved the highest Dice score of 68.07\% for D3. 
\begin{table}[!ht]
\centering
    \caption{Performance of wound tissue classification models using segmentation models.}
    \label{tab:ConvSeg}
% \resizebox{1\textwidth}{!}{
\begin{tabular}{clcccccc}
\hline
Dataset & Model & Precision (\%) & Recall (\%) & Dice (\%) & IoU (\%) & AUC & MCC (\%)\\
\hline \hline
\multirow{5}{*}{D1} & UNet & 47.29	&	28.59	&	35.25	&	24.68	&	0.92	&	32.83\\
   & UNet2D & \textbf{77.84}	&	68.36	&	\textbf{67.87}	&	\textbf{54.22}	&	\textbf{0.97}	&	\textbf{67.19}\\
   & UNet++ 2D & 75.87	&	\textbf{69.45}	&	65.56	&	51.5	&	\textbf{0.97}	&	65.88\\
   & R2UNet & 64.06	&	64.49	&	63.31	&	48.8	&	0.95	&	60.7\\
   & Attention 2DUNet & 60.45	&	68.02	&	62.3	&	51.83	&	0.96	&	60.83\\
   & cGAN	& 53.59	&	28.76	&	30.21	&	23.48	& -- &	31.77 \\
   \hline

\multirow{5}{*}{D3} & UNet & 47.36	&	61.91	&	44.44	&	34.42	&	0.91	&	45.01\\
   & UNet2D & 54.32	&	49.16	&	30.99	&	19.72	&	0.9	&	33.25\\
       & UNet++ 2D	& \textbf{64.63}	&	\textbf{81.17}	&	\textbf{68.07}	&	\textbf{54.44}	&	\textbf{0.96}	&	\textbf{67.08}\\
   & R2UNet	& 63.34	&	63.76	&	48.49	&	34.65	&	0.95	&	50.86\\
   & Attention 2DUNet	& 61.25	&	78.61	&	61.16	&	46.56	&	\textbf{0.96}	&	61.16\\
   & cGAN  & 54.07	&	8.17	&	12.45	&	7.12 & -- &	13.8 \\  
\hline

\hline
\end{tabular}
% }
\end{table}

\subsection{Performance of modified-segmentation models}   
Fifteen modified segmentation models were used for wound tissue segmentation across both the full image (D1) and superpixel (D3) forms of the dataset. These modified models are constructed from a combination of segmentation models (base models or decoders) and DL classification models (encoders). The segmentation models used included UNet, LinkNet, and Feature Pyramid Network (FPN), while the DL classification models consisted of VGG16, ResNet152, DenseNet201, InceptionV3, and MobileNetV2.

To evaluate the performance of the modified segmentation models, relevant evaluation metrics were calculated for all tissue classes and subsequently averaged based on class weight. The results are reported in Table \ref{tab:ModSeg}. The combination of FPN and VGG16 was observed to have a high Dice score and IoU of 76.95\% and 62.77\%, respectively, for D1. For D3, the highest Dice score and IoU of 52.68\% and 37.17\%, respectively, were obtained with the LinkNet-VGG16 combination. In terms of overall performance, the FPN-VGG16 model exhibited the highest Dice score of 76.95\%.
\begin{table}[!ht]
\centering
    \caption{Performance of wound tissue classification models using modified-segmentation models.}
    \label{tab:ModSeg}
% \resizebox{1\textwidth}{!}{
\begin{tabular}{cllcccccc}
\hline
Dataset & Model &
Backbone & Precision (\%) & Recall (\%) & Dice (\%) & IoU (\%) & AUC & MCC (\%)\\
\hline \hline
D1 & UNet & VGG16 & 57.78	&	58.37	&	57.05	&	41.2	&	0.91	&	54.03\\
   &     & ResNet152 & 65.59	&	60.15	&	55.54	&	38.77	&	0.86	&	56.79\\
   &     & DenseNet201 	& 68.2	&	67.79	&	66.68	&	52.3	&	0.88	&	65.5\\
   &     & InceptionV3 	&  62.73	&	69.97	&	62.09	&	47.97	&	0.89	&	61.9\\
   &     & MobileNetV2	&  23.5	&	25.69	&	23.85	&	14.16	&	0.57	&	10.15\\
   \cline{2-9}
   
   & LinkNet & VGG16	&  57.54	&	61.77	&	58.21	&	42.65	&	0.87	&	55.33\\
   &         & ResNet152 	&  75.24	&	54.89	&	57.39	&	41.83	&	0.91	&	58.43\\
   &         & DenseNet201	&  20.7	&	27.56	&	22.11	&	13.08	&	0.55	&	17.47\\
   &         & InceptionV3 	&  45.5	&	53.43	&	46.28	&	31.47	&	0.77	&	44.46\\
   &         & MobileNetV2 	&  13.91	&	33.22	&	17.77	&	10.42	&	0.61	&	13.89\\
  \cline{2-9}
   
   & FPN     & VGG16 	& \textbf{83.47}	&	\textbf{72.34}	&	\textbf{76.95}	&	\textbf{62.77}	&	\textbf{0.98}	&	\textbf{75.87}\\
   &         & ResNet152	& 41.09	&	56.83	&	46.5	&	31.34	&	0.93	&	43.71\\
   &         & DenseNet201	& 61.12	&	43.45	&	46.66	&	32.51	&	0.91	&	47.05\\
   &         & InceptionV3	& 66.25	&	60.46	&	63.17	&	46.94	&	0.96	&	61.01\\
   &         & MobileNetV2 	& 52.67	&	61.27	&	53.09	&	38.22	&	0.94	&	52.18\\
  \hline

D3 & UNet & VGG16 	& 54.82	&	52.05	&	50.85	&	35.44	&	0.86	&	48.37\\
   &     & ResNet152	&  \textbf{62.53}	&	\textbf{59.12}	&	49.21	&	36.17	&	0.88	&	\textbf{51.08}\\
   &     & DenseNet201 	&  33.96	&	28.35	&	26.77	&	15.77	&	0.81	&	24.48\\
   &     & InceptionV3	&  52.95	&	56.72	&	47.66	&	32.12	&	0.83	&	47.58\\
   &     & MobileNetV2 	& 45.8	&	39.83	&	21.75	&	13.65	&	0.86	&	22.08\\
  \cline{2-9}
   
   & LinkNet & VGG16	& 54.58	&	54.93	&	\textbf{52.68}	&	\textbf{37.17}	&	\textbf{0.92}	&	49.99\\
   &         & ResNet152 	& 50.26	&	36.22	&	38.42	&	27.31	&	0.84	&	37.87\\
   &         & DenseNet201	&  55.69	&	59.01	&	41.94	&	28.76	&	0.88	&	44.65\\
   &         & InceptionV3	&  44.11	&	43.03	&	30.79	&	18.69	&	0.75	&	32.04\\
   &         & MobileNetV2 	&  56.59	&	46.71	&	43.65	&	28.08	&	\textbf{0.92}	&	44\\
   \cline{2-9}
   
   & FPN     & VGG16 	& 51.16	&	28.45	&	36.47	&	22.76	&	0.88	&	34.97\\
   &         & ResNet152 	& 59.74	&	22.86	&	28.31	&	19.27	&	0.55	&	29.28\\
   &         & DenseNet201	&  21.44	&	17.02	&	16.69	&	10.27	&	0.68	&	14.42\\
   &         & InceptionV3 	& 49.5	&	37.74	&	31.61	&	19.64	&	0.87	&	32.82\\
   &         & MobileNetV2 	&  43.01	&	30.79	&	20.91	&	13.5	&	0.9	&	21.86\\

\hline

\hline
\end{tabular}
% }
\end{table}

\subsection{Performance of DL-based classification models} 
To evaluate the DL models for tissue classification, we tested eight DL architectures using the D2 and D3 dataset forms. The models included VGG16, ResNet50, DenseNet201, EfficientNetB7, MobileNetV2, InceptionV3, NASNetMobile, and Xception. The results are presented in Table \ref{tab:DLClass}.

It was observed that NASNetMobile achieved a high F1 score of 77.13\% for D2, while DenseNet201 attained the highest F1 score of 61.39\% for D3. Additionally, the best Matthews correlation coefficient (MCC) score of 65.97\% was found for NASNetMobile using D2, and 46.4\% for DenseNet201 using D3. In terms of overall performance, NASNetMobile exhibited the highest MCC score of 65.97\%. 
\begin{table}[!ht]
\centering
    \caption{Performance of wound tissue classification models using DL-based approaches.}
    \label{tab:DLClass}
\resizebox{1\textwidth}{!}{
\begin{tabular}{clccccccc}
\hline
Dataset & Model & Precision (\%) & Recall (\%) & Specificity (\%) & F1 Score (\%) & IoU (\%) & AUC & MCC (\%)\\
\hline \hline
D2 & VGG16 	& 52.44	&	38.06	&	85.63	&	40.63	&	27.51	&	0.76	&	26\\
   & ResNet50	& 78.43	&	72.82	&	88.88	&	\textbf{74.13}	&	\textbf{59.59}	&	0.91	&	65.15\\
   & DenseNet201 & \textbf{79.42}	&	72.35	&	85.93	&	71.37	&	56.19	&	0.93	&	63.82\\
   & EfficientNetB7 	& 53.11	&	35.66	&	82.39	&	28.66	&	18.17	&	0.84	&	19.95\\
   & MobileNetV2  & 73.03	&	60.07	&	88.55	&	62.53	&	46.16	&	0.88	&	52.39\\
   & InceptionV3 & 70.18	&	58.76	&	87.01	&	61.75	&	44.71	&	0.86	&	49.2\\
   & NASNetMobile  & 77.52	&	\textbf{74.68}	&	88	&	\textbf{74.13}	&	59.3	&	\textbf{0.94}	&	\textbf{65.97}\\
   & Xception 	& 64.93	&	42.08	&	\textbf{90.15}	&	44.86	&	30.56	&	0.83	&	36.81\\
   \hline
   
D3 & VGG16	& 52.34	& 49.97	& 77.48	& 49.42	& 34.73	& 0.71	& 28.19\\
   & ResNet50 	& 57.64	& 58.29	& 79.35	& 56.76	& 42.18	& 0.77	& 38.80\\
   & DenseNet201 & \textbf{63.75}	& \textbf{65.26}	& 79.56	& \textbf{61.39}	& \textbf{47.25}	& \textbf{0.80}	& \textbf{46.4}\\
   & EfficientNetB7 	& 31.62	& 50.24	& 61.41	& 37.41	& 27.21	& 0.63	& 17.13\\
   & MobileNetV2  & 59.27	& 57.55	& \textbf{82.33}	& 56.68	& 41.72	& 0.71	& 40.12\\
   & InceptionV3 	& 54.9	& 54.83	& 78.36	& 52.83	& 38.38	& 0.70	& 32.94\\
   & NASNetMobile 	& 50.18	& 53.31	& 72.09	& 49.82	& 35.29	& 0.73	& 26.82\\
   & Xception	& 51.18	& 50.56	& 75.63	& 50.2	& 34.89	& 0.69	& 26.86\\

\hline

\hline
\end{tabular}
}
\end{table}

\subsection{Performance of modified-classification approaches}
In the modified-classification approaches, DL classification models are used as feature extractors and are combined with traditional ML models as classifiers for wound tissue classification. We used twelve modified-classification approach-based models, where the DL models were AlexNet, VGG16, InceptionV3, and DenseNet201, and the ML models included SVM, RF, and KNN. 

Table \ref{tab:DLMLClass} reports the performance of these models. The results indicate that the combination of AlexNet and RF showed high performance across all evaluation metrics for D2. For D3, the combination of DenseNet201 and RF emerged as the best model, based on the F1 and MCC scores. Overall, the combination of AlexNet and RF achieved the highest F1 score and MCC of 76.19\% and 67.88\%, respectively.
\begin{table}[!ht]
\centering
    \caption{Performance of wound tissue classification using modified-classification approaches.}
    \label{tab:DLMLClass}
\resizebox{1\textwidth}{!}{
\begin{tabular}{clcccccccc}
\hline
Dataset & DL Feature &
ML Model & Precision (\%) & Recall (\%) & Specificity (\%) & F1 Score (\%) & IoU (\%) & AUC & MCC (\%)\\
\hline \hline
D2 & AlexNet & SVM 	& 74.35	&	62.09	&	90.9	&	66.45	&	50.88	&	0.9	&	56.02\\
   &         & RF 	& \textbf{80.7}	&	\textbf{73.38}	&	\textbf{92.52}	&	\textbf{76.19}	&	\textbf{62.03}	&	\textbf{0.92}	&	\textbf{67.88}\\
   &         & KNN 	& 76.82	&	68.38	&	90.98	&	71.91	&	56.82	&	0.86	&	61.83\\
   \cline{2-10}
   & VGG16 & SVM 	& 69.05	&	55.73	&	90.73	&	59.37	&	45.13	&	0.87	&	48.99\\
   &         & RF 	& 68.22	&	58.49	&	88.37	&	60.61	&	45.46	&	0.86	&	48.97\\
   &         & KNN 	& 67.94	&	57.78	&	89.3	&	59.73	&	45.19	&	0.77	&	48.9\\
  \cline{2-10}
  
   & InceptionV3 & SVM & 69.92	&	56.62	&	89.89	&	60.58	&	44.96	&	0.9	&	49.49\\
   &              & RF 	& 74.21	&	62.94	&	89.84	&	65.81	&	49.59	&	0.9	&	56.15\\
   &              & KNN & 66.31	&	48.59	&	87.68	&	52.23	&	36.07	&	0.78	&	40.62\\
  \cline{2-10}
  
   & DenseNet201 & SVM 	&69.99	&	58.32	&	89.99	&	62.75	&	48.83	&	0.87	&	51.03\\
   &             & RF 	& 77.96	&	68.02	&	92.15	&	71.67	&	57.64	&	0.9	&	62.42\\
   &             & KNN 	& 76.19	&	66.71	&	91.55	&	70.45	&	56.24	&	0.85	&	60.47\\
\hline

D3 & AlexNet & SVM 	& 52.68	& 53.03	& 70.87	& 47.95	& 34.05	& 0.71	& 27.13\\
   &         & RF	& 50.57	& 49.32	& 65.1	& 41.84	& 29.28	& 0.65	& 19.9\\
   &         & KNN	& 50.05	& 50	& 73.54	& 47.99	& 33.46	& 0.66	& 25.13\\
   \cline{2-10}
   & VGG16 & SVM 	& 56.08	& 53.52	& 78.12	& 52.8	& 37.77	& 0.79	& 33.14\\
   &         & RF 	& 57.4	& 57.51	& 78.83	& 56.54	& 41.24	& 0.79	& 37.28\\
   &         & KNN 	& 53.73	& 51.59	& 77.55	& 50.64	& 36	& 0.72	& 30.24\\
  \cline{2-10}
  
   & InceptionV3 & SVM	& 54.67	& 52.98	& 77.25	& 52.69	& 37.55	& 0.70	&31.49\\
   &              & RF 	& 54.41	& 55.17	& 74.43	& 52.29	& 37.85	& 0.73	& 32.3\\
   &              & KNN 	& 53.64	& 52.24	& 74.69	& 51.05	& 35.99	& 0.67	& 28.97\\
  \cline{2-10}
  
   & DenseNet201 & SVM 	& 63.4	& 60.92	& 82.56	& 61.12	& 45.84	& \textbf{0.81}	& 44.72\\
   &             & RF 	& \textbf{63.58}	& \textbf{61.7}	& \textbf{82.68}	& \textbf{61.42}	& \textbf{46.64}	& 0.79	& \textbf{45.67}\\
   &             & KNN 	& 59.02	& 56.05	& 81.5	& 56.25	& 41.27	& 0.73	& 38.55\\

\hline

\hline
\end{tabular}
}
\end{table}

\subsection{Applying class weight}
Our dataset exhibits imbalanced classes. To mitigate this class imbalance, we used class weights, assigning weights to classes based on their sizes. The primary purpose of using class weights is to penalise the misclassification of minority classes (such as bone and tendon) by assigning them higher weights, while simultaneously reducing the weights for majority classes (such as granulation, slough, maceration, and necrosis). 

Table \ref{tab:Classweight} reports the results of this approach. They indicate that the class weight approach did not yield significant improvements compared to the results obtained without class weights (Table \ref{tab:DLClass}) for both the D2 and D3 datasets. Specifically, the highest MCC score was 66.37\% for D2 and 46.14\% for D3 with class weights, while the highest MCC score without class weights was 65.97\% for D2 and 46.4\% for D3. A similar pattern was observed for the F1 score.    
\begin{table}[!ht]
\centering
    \caption{Performance of wound tissue classification models using DL-based approaches by applying class weights.}
    \label{tab:Classweight}
\resizebox{1\textwidth}{!}{
\begin{tabular}{clccccccc}
\hline
Dataset & Model & Precision (\%) & Recall (\%) & Specificity (\%) & F1 Score (\%) & IoU (\%) & AUC & MCC (\%)\\
\hline \hline
D2 & VGG16  &	59.4	&	46.74	&	79.06	&	46.22	&	30.5	&	0.79	&	30.05\\
   & ResNet50  &	78.5	&	\textbf{74.95}	&	88.55	&	\textbf{74.36}	&	\textbf{60.77}	&	\textbf{0.91}	&	\textbf{66.37}\\
   & DenseNet201  &	73.42	&	53.78	&	86.18	&	59.56	&	43.2	&	0.83	&	46.93\\
   & EfficientNetB7 &	44.26	&	34.96	&	96.29	&	37.55	&	28.48	&	0.78	&	34.28\\
   & MobileNetV2  &	67.02	&	32.63	&	91.46	&	42.74	&	28.15	&	0.81	&	32.4\\
   & InceptionV3  & 58.99	&	36.47	&	85.66	&	44.51	&	29.7	&	0.74	&	28.17\\
   & NASNetMobile &	\textbf{81}	&	46.23	&	\textbf{96.96}	&	48.79	&	39.58	&	0.82	&	48.2\\
   & Xception  &	61.97	&	35.78	&	87.23	&	42.1	&	27.01	&	0.79	&	28.96\\
   \hline
   
D3 & VGG16	& 54.22 & 42.29 & 83.41 &	45.66 & 30.36 &	0.73 &	26.41\\
   & ResNet50  &	64.19 &	55.35 &	85.14 &	57.05 &	41.59 &	\textbf{0.77} &	41.52\\
   & DenseNet201  &	\textbf{66.64} &	\textbf{57.69} &	\textbf{88.41} &	\textbf{59.20} &	\textbf{43.56} &	\textbf{0.77} &	\textbf{46.14}\\
   & EfficientNetB7  &	43.18 &	50.33 &	58.11 &	37.93 &	27.06 &	0.63 &	17.15\\
   & MobileNetV2  &	58.86 &	52.50 &	85.12 &	55.15 &	40.98 &	0.72 &	38.33\\
   & InceptionV3  &	56.29 &	49.90 &	82.73 &	51.83 &	36.03 &	0.71 &	32.61\\
   & NASNetMobile  &	54.55 &	40.18 &	84.86 &	43.83 &	29.1 &	0.74 &	25.79\\
   & Xception  &	52.28 &	39.72 &	82.46 &	43.91 &	29.16 &	0.69 &	23.49\\
   \hline

\hline

\hline
\end{tabular}
}
\end{table}

\subsection{Tissue-wise performance of segmentation and classification models}\label{sec:tissuewise}
Table~\ref{tab:TissueWise} presents the tissue-wise performance for the selected segmentation and classification models across three dataset forms: D1 (full image), D2 (patching), and D3 (superpixel). We selected the two best-performing models for each form of the dataset applied to the segmentation and classification approaches. The results show that segmentation models, particularly in the D1 dataset, excelled at identifying granulation and necrotic tissues. Notably, the FPN+VGG16 model demonstrated superior performance, achieving high Dice scores of 89.6\% and 78.32\% for granulation and necrosis, respectively. However, across all dataset forms, bone and tendon tissues posed significant challenges, with both UNet++ 2D and LinkNet+VGG16 in the D3 dataset failing to segment these tissues, as reflected by 0\% Dice scores for bone and very low scores for tendon.
\begin{table}[!htb]
\centering
    \caption{Performance of segmentation and classification approaches for tissue-wise classification}
    \label{tab:TissueWise}
\resizebox{.9\textwidth}{!}{
\begin{tabular}{ccllcccccc}
\hline
\multicolumn{1}{c}{\multirow{2}{*}{Approach}} & \multicolumn{1}{c}{\multirow{2}{*}{Dataset}} & \multicolumn{1}{c}{\multirow{2}{*}{Model}} &
\multicolumn{1}{l}{\multirow{2}{*}{Metrics}} & \multicolumn{6}{c}{Tissue classes} \\ \cline{5-10}
\multicolumn{1}{c}{} & \multicolumn{1}{c}{} & \multicolumn{1}{c}{} & \multicolumn{1}{c}{} & Slough & Granulation & Bone & Tendon & Maceration & Necrosis\\
\hline \hline
\multirow{20}{*}{\rotatebox[origin=c]{90}{Segmentation}} & \multirow{10}{*}{D1} & UNet2D & Precision (\%) & 35.65	&	93.28	&	24.81	&	17.33	&	83.45	&	92.82  \\
  & &     & Recall (\%) & 84.36	&	86.43	&	5.5	&	3.26	&	24.85	&	64.49\\
  & &     & Dice (\%) & 50.12	&	89.72	&	9.01	&	5.49	&	38.3	&	76.1\\
  & &     & IoU (\%)	&  33.44	&	81.36	&	4.72	&	2.82	&	23.68	&	61.42\\
  & &     & MCC (\%)	& 49.5	&	86.99	&	11.49	&	7.09	&	42.78	&	77.02\\
   \cline{3-10}
   
&  & FPN+VGG16 & Precision (\%)  & 73.32	&	88.12	&	83.61	&	51.37	&	90.18	&	94.55\\
 &  &     & Recall (\%) & 71.76	&	91.14	&	56.15	&	95.35	&	76.5	&	66.84\\
 &  &     & Dice (\%) & 72.53	&	89.6	&	67.18	&	66.77	&	82.78	&	78.32\\
 &  &     & IoU (\%)	&  56.9	&	81.16	&	50.58	&	50.12	&	70.62	&	64.36\\
 &  &     & MCC (\%)	&  70.28	&	86.55	&	68.4	&	69.59	&	81.29	&	79.18\\
 
    \cline{2-10}
 
 & \multirow{10}{*}{D3} & UNet++ 2D & Precision (\%) &  35.31	&	92.92	&	0	&	0	&	81.4	&	80.52\\
  & &     & Recall (\%) &  85.92	&	84.24	&	0	&	0	&	23.48	&	76.9\\
  & &     & Dice (\%) & 50.05	&	88.37	&	0	&	0	&	36.45	&	78.67\\
  & &     & IoU (\%)	&  33.38	&	79.16	&	0	&	0	&	22.29	&	64.84\\
  & &     & MCC (\%)	& 49.75	&	85.37	&	0	&	0	&	40.9	&	78.28\\
   \cline{3-10}
   
&  & LinkNet+VGG16 & Precision (\%)  & 35.96	&	92.35	&	0	&	12.35	&	67.41	&	72.91\\
 &  &     & Recall (\%) & 55.14	&	87.6	&	0	&	14.59	&	37.46	&	75.93\\
 &  &     & Dice (\%) & 43.53	&	89.91	&	0	&	13.37	&	48.16	&	74.39\\
 &  &     & IoU (\%)	&  27.82	&	81.67	&	0	&	7.17	&	31.71	&	59.22\\
 &  &     & MCC (\%)	&  38.8	&	87.14	&	-0.09	&	12.39	&	46.25	&	73.89\\
 
    \cline{1-10} 
 
 \multirow{20}{*}{\rotatebox[origin=c]{90}{Classification}} & \multirow{10}{*}{D2} & NASNetMobile & Precision (\%) & 68	&	74	&	0	&	100	&	52	&	98\\
   &  &   & Recall (\%) & 93	&	66	&	0	&	1	&	34	&	60\\
   &   &  & F1 Score (\%) & 79	&	70	&	0	&	2	&	41	&	75\\
   &  &    & IoU (\%)	&  65.24	&	53.84	&	0	&	0.81	&	25.74	&	59.62\\
   &   &  & MCC (\%)	& 66.8	&	61.51	&	-0.05	&	8.96	&	41.16	&	72.59\\
   \cline{3-10}
   
&  & AlexNet+RF & Precision (\%)  & 78	&	71	&	0	&	2	&	53	&	99\\
   &  &   & Recall (\%) & 65	&	90	&	0	&	1	&	28	&	74\\
   &  &   & F1 Score (\%) & 71	&	80	&	0	&	1	&	36	&	85\\
   &  &   & IoU (\%)	&  54.88	&	66.01	&	0	&	0.56	&	22.14	&	73.35\\
   &  &   & MCC (\%)	&  57.66	&	73.02	&	-0.99	&	0.96	&	37.51	&	82.23\\
 
    \cline{2-10}
 
 & \multirow{10}{*}{D3} & DenseNet201 & Precision (\%) & 58	&	70	&	0	&	67	&	62	&	36\\
  & &     & Recall (\%) & 14	&	86	&	0	&	27	&	70	&	57\\
  & &     & F1 Score (\%) & 23	&	77	&	0	&	38	&	66	&	44\\
  & &     & IoU (\%)	&  13.04	&	62.76	&	0	&	23.53	&	49.12	&	28.57\\
  & &     & MCC (\%)	& 22.44	&	54.36	&	0	&	41.41	&	50.8	&	43.46\\
   \cline{3-10}
   
&  & DenseNet201+RF & Precision (\%)  & 34	&	74	&	0	&	0	&	76	&	27\\
 &  &     & Recall (\%) & 53	&	80	&	0	&	0	&	49	&	19\\
 &  &     & F1 Score (\%) & 41	&	77	&	0	&	0	&	60	&	22\\
 &  &     & IoU (\%)	&  25.78	&	62.56	&	0	&	0	&	42.86	&	12.5\\
 &  &     & MCC (\%)	&  25.09	&	55.75	&	0	&	-0.57	&	49.98	&	20.48\\

\hline

\hline
\end{tabular}
}
\end{table}

In classification tasks, models trained on the D2 dataset showed better performance for granulation and necrotic tissues, with NASNetMobile and AlexNet+RF achieving F1 scores of 70\% and 85\% for granulation and necrosis, respectively. However, the classification of bone and tendon remained problematic across all datasets, with F1 scores dropping to 0\% for bone and minimal scores for tendon. The poor performance of both segmentation and classification models for bone and tendon can be attributed to the severe class imbalance, where these tissues are heavily underrepresented, particularly in the D3 dataset.

To assess the impact of these underrepresented tissue classes on the overall segmentation performance, we designed an experiment where we ran the best two performing models for the D1 dataset, UNet 2D and FPN+VGG16, exclusively on the slough, granulation, maceration, and necrosis classes, while excluding the bone and tendon classes. The rationale behind this experiment was to determine how the poor representation and segmentation of bone and tendon in Table~\ref{tab:TissueWise} might have influenced the models' overall performance. The results of this experiment, presented in Table~\ref{tab:FourTissueClass}, demonstrate a clear improvement in segmentation performance across the remaining tissue classes when bone and tendon were removed.
\begin{table}[!htbp]
\centering
    \caption{Performance of segmentation approaches for slough, granulation, maceration, and necrotic tissue classes.}
    \label{tab:FourTissueClass}
\resizebox{1\textwidth}{!}{
\begin{tabular}{cllccccc}
\hline
\multirow{2}{*}{Dataset} & \multirow{2}{*}{Model} &
\multirow{2}{*}{Metrics} & \multicolumn{4}{c}{Tissue type} & \multirow{2}{*}{Weighted Avg.}\\ \cline{4-7}
 & &  & Slough & Granulation & Maceration & Necrosis &\\
\hline \hline
D1 & UNet 2D & Precision (\%) & 42.29	&	94	&	81.41	&	85.62	&	70.25\\
   &     & Recall (\%) & 76.52	&	87.14	&	47.44	&	68.38	&	76.61\\
   &     & Dice (\%) & 54.48	&	90.44	&	59.95	&	76.03	&	70.98\\
   &     & IoU (\%)	&  37.44	&	82.55	&	42.81	&	61.33	&	57.26\\
   &     & MCC (\%)	&  52.22	&	87.91	&	59.08	&	76.11	&	69.3\\
   \cline{2-8}
   
   & FPN+VGG16 & Precision (\%)  & 74.99	&	89.32	&	90.95	&	88.62	&	83.33\\
   &     & Recall (\%) & 75.04	&	93.49	&	82.2	&	57.5	&	76.09\\
   &     & Dice (\%) & 75.02	&	91.36	&	86.35	&	69.75	&	78.82\\
   &     & IoU (\%)	&  60.02	&	84.09	&	75.98	&	53.55	&	65.93\\
   &     & MCC (\%)	&  72.94	&	88.82	&	84.99	&	70.96	&	77.49\\

\hline

\hline
\end{tabular}
}
\end{table}

For the UNet 2D model, excluding bone and tendon led to a marked increase in performance. In Table~\ref{tab:TissueWise}, the model struggled to segment bone and tendon, yielding very low Dice scores of 9.01\% and 5.49\%, respectively. However, after removing these challenging tissue classes, the weighted average Dice score increased significantly from 67.87\% (Table~\ref{tab:ConvSeg}) to 70.98\%. This improvement was observed across all remaining tissue classes. For instance, the Dice score for granulation increased slightly from 89.72\% (Table~\ref{tab:TissueWise}) to 90.44\%, while the performance for maceration showed a notable rise from 38.3\% to 59.95\%. This indicates that bone and tendon had been negatively impacting the model's overall performance, and once excluded, the model's ability to focus on better-represented classes improved considerably.

The FPN+VGG16 model, which was already performing well (see in Table~\ref{tab:TissueWise}), further benefited from the exclusion of bone and tendon. Its weighted average Dice score rose from 76.95\% (Table~\ref{tab:ModSeg}) to 78.82\%, reflecting enhanced performance across all remaining tissue types. For example, the granulation Dice score improved from 89.6\% to 91.36\%, and the maceration Dice score increased significantly from 82.78\% to 86.35\%. Although there was a slight drop in the necrosis Dice score from 78.32\% to 69.75\%, the overall improvement in performance highlights the model's capacity to segment well-represented tissues more effectively when not hampered by the underrepresented and poorly segmented bone and tendon classes. Despite all these challenges, the D1 form of the dataset, when paired with the FPN+VGG16 model, demonstrated the best overall performance for wound tissue classification.

\subsection{Visual performance analysis}
\begin{figure}[!tb]
    \centering
    \includegraphics[width=1\textwidth]{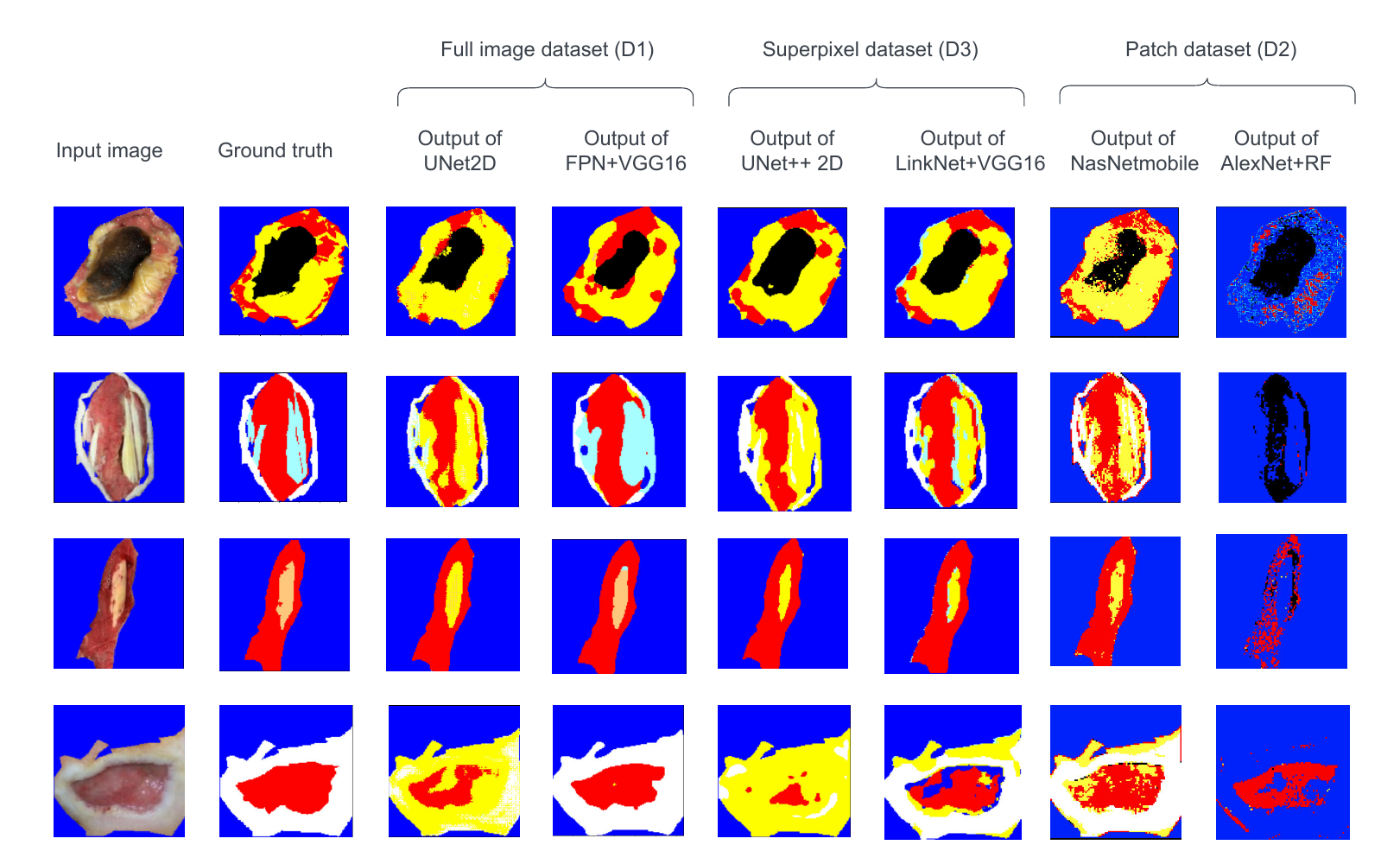}
    \caption{Visual comparison of predicted wound tissue segmentation using two best-performing models across three forms, D1, D2, and D3, of the dataset. The first column from the left shows input images, while the second column illustrates the ground truth, where different tissue types are colour-coded: granulation in red, slough in yellow, maceration in white, necrosis in black, tendon in sky blue, and bone in cream. The remaining columns display the predicted segmentation results from the two best-performing models for each dataset form, showcasing the accuracy and limitations of each model in identifying and segmenting the wound tissues.}
    \label{fig:ModelPredicted}
\end{figure}

Figure~\ref{fig:ModelPredicted} provides a visual comparison of the best-performing models' predicted images alongside the corresponding input and ground truth images for the three dataset forms: D1 (full image), D2 (patch), and D3 (superpixel). This visual performance analysis highlights the strengths and limitations of various segmentation and classification models in accurately identifying wound tissues.

For D1, the FPN+VGG16 segmentation model demonstrated superior performance by accurately segmenting most wound tissues. The clear delineation of tissue boundaries in the predicted images closely resembles the ground truth, indicating the model's effectiveness in identifying complex tissue structures, such as granulation and necrosis. UNet 2D, however, struggled with more intricate tissue types, particularly bone and tendon. This conventional model failed to correctly classify these tissues, often misclassifying them as slough, which aligns with its poor performance metrics reported earlier. This visual evidence underscores the limitations of standard segmentation models in handling underrepresented tissue classes like bone and tendon, suggesting an opportunity for future model development in this area.

In D3, the LinkNet+VGG16 model performed better than the UNet++ 2D, demonstrating strong segmentation capabilities for granulation and necrotic tissues. Despite the lower representation of certain tissue types in the D3 form of the dataset, this model could still identify the major wound tissues with reasonable accuracy. In contrast, UNet++ 2D again exhibited poor segmentation results, particularly in identifying bone and tendon, which were either missed entirely or often classified inaccurately as slough. This suggests that while superpixel-based datasets offer a more compact representation, they might oversimplify the visual features required to accurately segment complex tissue types.

For D2, NASNetMobile demonstrated considerable success in classifying wound tissues, showing strong recognition of granulation, necrosis, and other major tissue types, except for bone and tendon. The visual results show that the model could identify the overall structure of the wound, but finer details, such as blonde and tendon segmentation, were still problematic. This indicates that while NASNetMobile is well-suited for broader tissue classification tasks, it struggles with highly localised and underrepresented classes. On the other hand, the combination of AlexNet with random forest (RF) produced subpar results, as seen in the visual outputs. The predicted images from this model could not correctly identify most wound tissues, resulting in a significant deviation from the ground truth. This highlights the limitations of combining traditional machine learning models with deep learning architectures, particularly for tasks requiring high precision and spatial understanding.

Overall, the visual analysis in Figure~\ref{fig:ModelPredicted} reinforces the earlier quantitative findings. While FPN+VGG16 emerges as the most effective model for wound tissue segmentation on the D1 form of the dataset, conventional models like UNet2D and UNet++ 2D lag, particularly in classifying bone and tendon tissues. Regarding classification, NASNetMobile performs well for D2 but struggles with more complex tissue types, highlighting the need for further refinement in models to handle underrepresented and difficult-to-segment tissues. These findings suggest a clear direction for future research, focusing on improving model architectures for bone and tendon segmentation and exploring more sophisticated approaches for handling imbalanced datasets.

\subsection{Leave-one-out cross-validation}
This study used leave-one-out cross-validation (LOOCV) on the D1 form of the dataset to evaluate the robustness of deep learning (DL) models for wound tissue segmentation. LOOCV is particularly valuable for estimating accurate model performance in studies with smaller datasets \citep{RT4sammut2010leave}. We tested the best-performing segmentation model, FPN+VGG16, using LOOCV.

In LOOCV, the number of folds is equal to the total number of images in the dataset, which in our case was 187. Consequently, the experiment was run 187 times, with each iteration using a different single image for testing and the remaining images for training and validation (with 1\% of the training set allocated for validation). The final results were calculated as a weighted average over 187 iterations, using the number of pixels as the weight. To assess the effect of class weighting on our highly imbalanced dataset, we conducted LOOCV both with and without applying class weights to categorical cross-entropy as the loss function. 

\begin{figure}[htbp]
    \centering
    \begin{tikzpicture}
    \begin{axis}[
        width  = 0.85*\textwidth,
        height = 8cm,
        major x tick style = transparent,
        ybar=2*\pgflinewidth,
        bar width=14pt,
        ymajorgrids = false,
        ylabel = {Dice score (\%)}, % y-axis label
        symbolic x coords={Slough,Granulation,Bone,Tendon,Maceration,Necrosis, Weighted Avg.},
        xtick = data,
        scaled y ticks = false,
        enlarge x limits=0.10,
        ymin=0, ymax=115,
        legend cell align=left,
        nodes near coords,
        every node near coord/.append style={rotate=90,anchor=west},
        legend style={
                at={(.5,1.05)},
                anchor=south,legend columns=-1, column sep=1ex
        }
    ]
        % % Split+w/o class weight
        % \addplot[style={orange,fill=orange,mark=none}] coordinates {
        %     (Slough, 72.53) 
        %     (Granulation, 89.60) 
        %     (Bone, 67.18) 
        %     (Tendon, 66.77) 
        %     (Maceration, 82.78) 
        %     (Necrosis, 78.32) 
        %     (Average, 76.95)
        % };
        
        % LOOCV+w/0 classweight
        \addplot+[pattern=dots,pattern color=blue] coordinates {
        (Slough,78.00) 
        (Granulation,88.39)
        (Bone,24.28) 
        (Tendon,31.76) 
        (Maceration,77.06) 
        (Necrosis,94.37) 
        (Weighted Avg.,82.25)
        };

        % LOOCV+with class weight
        \addplot+[pattern=north east lines, pattern color=orange] coordinates {
            (Slough, 72.16) 
            (Granulation, 81.30) 
            (Bone, 42.94) 
            (Tendon, 32.55) 
            (Maceration, 82.48) 
            (Necrosis, 94.05) 
            (Weighted Avg., 78.31)
        };
\legend{Without class-weight, With class-weight}
\end{axis}
    \end{tikzpicture}
    \caption{Impact of using class-weight on the segmentation performance of the FPN+VGG16 model evaluated using leave-one-out cross validation}
    \label{fig:loocv-class-weight-impact}
\end{figure}
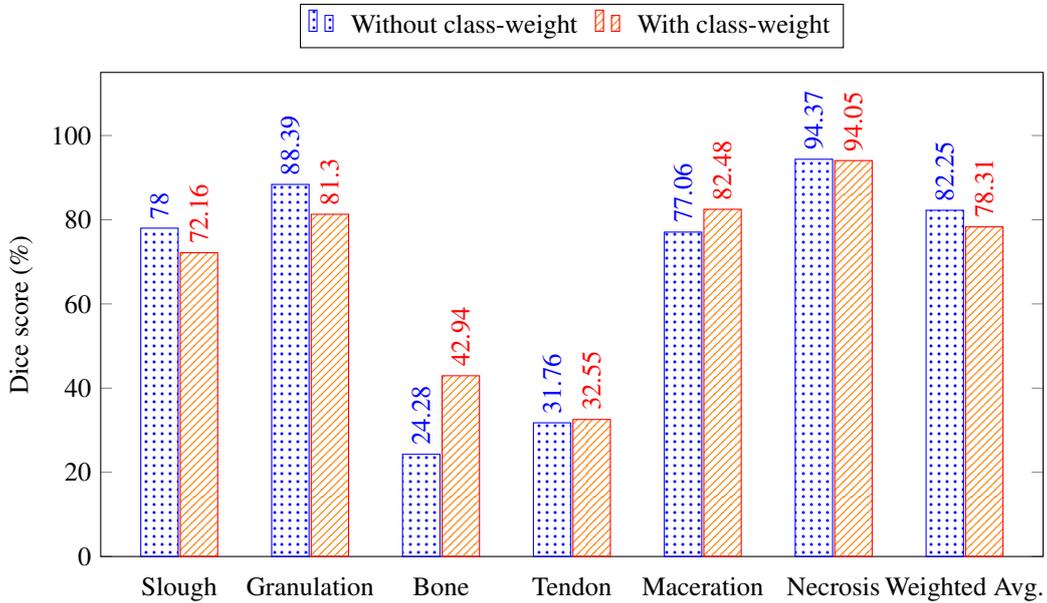

Table~\ref{tab:leaveoneout} summarises the performance of the FPN+VGG16 model using LOOCV for various tissue types. The weighted average values across all tissue types indicate the model's general performance. Without class weighting, the weighted average Dice score was 82.25\%, and IoU was 75.45\%. With class weighting, these metrics decreased slightly, with the Dice score averaging 78.31\%, and IoU 70.14\%. Figure~\ref{fig:loocv-class-weight-impact} directly compares how class weighting influences model performance (Dice score), particularly for underrepresented tissues like bone and tendon. With class weighting applied, the Dice score for bone increases significantly from 24.28\% (without class weighting) to 42.94\%, and there is a slight improvement in the tendon from 31.76\% to 32.55\%, and in maceration from 77.06\% to 82.48\%. Although this shows a modest improvement in segmentation for these classes, the Dice scores for well-represented tissues like granulation, slough, and necrosis decrease a little. These results provide valuable insights into the trade-offs between class-weighted and non-weighted training approaches. Class weighting improves segmentation accuracy for underrepresented tissues, particularly bone and tendon, which have historically low segmentation performance due to their limited representation in the dataset. However, this improvement comes at the expense of reducing overall accuracy for well-represented classes, as the model must allocate more learning capacity to the difficult-to-segment tissues, which detracts from performance on tissues like granulation and necrosis.
\begin{table}[!htb]
\centering
    \caption{Performance of FPN+VGG16 model evaluated using leave-one-out cross-validation}
    \label{tab:leaveoneout}
\resizebox{1\textwidth}{!}{
\begin{tabular}{cp{3cm}lccccccc}
\hline
\multirow{2}{*}{Dataset} & \multirow{2}{*}{Model} &
\multirow{2}{*}{Metrics} & \multicolumn{6}{c}{Tissue type/condition} & \multirow{2}{*}{Weighted Avg.}\\ \cline{4-9}
&  & & Slough & Granulation & Bone & Tendon & Maceration & Necrosis & \\
\hline \hline
\multirow{10}{*}{D1} & \multirow{5}{2.8cm}{FPN+VGG16 (w/o class weight)} & Precision (\%) & 86.44 & 95.19 & 62.35 & 70.69 & 90.66 & 96.21 & 91.28\\
   &     & Recall (\%) & 75.68	& 85.20 & 17.31 & 28.26 & 74.61 & 94.01 & 79.58\\
   &     & Dice (\%) & 78.00 & 88.39 &	24.28 & 31.76 & 77.06 & 94.37 & \textbf{82.25}\\
   &     & IoU (\%)	& 71.00	& 81.83 & 16.41 & 23.52	& 68.40 & 90.47 & 75.45\\
   &     & MCC (\%)	& 75.51	& 85.54	& 28.51 & 33.58 & 71.87 & 93.04	& 79.51\\
   \cline{2-10}
   
   & \multirow{5}{2.8cm}{FPN+VGG16 (with class weight)} & Precision (\%)  & 79.57	&	94.12	&	45.71	&	64.59	&	83.50	&	91.99	&	83.50\\
   &     & Recall (\%) & 71.48	& 75.90	& 47.71	& 36.02	& 85.83	& 96.66	& 76.41\\
   &     & Dice (\%) & 72.16	& 81.30	& \textbf{42.94}	& \textbf{32.55}	& 82.48	& 94.05	& 78.31\\
   &     & IoU (\%)	&  63.80 &	72.84	& 31.67 &  22.81 &	74.25 &	89.18 &	70.14\\
   &     & MCC (\%)	&  67.69 & 78.16 & 43.09 & 34.56 & 74.41 & 91.64 & 74.40 \\

\hline

\hline
\end{tabular}
}
\end{table}

\section{Discussion}

\subsection{Findings and observations}

\paragraph{Model architecture} The performance discrepancies between UNet++, LinkNet+VGG16, and FPN+VGG16 underscore the importance of model architecture in tissue classification tasks. While UNet++ and LinkNet+VGG16 struggle with certain tissues like tendons and bones, FPN+VGG16 demonstrates superior performance, particularly with multi-scale feature recognition. This suggests that FPN's pyramid structure is more adept at capturing the hierarchical features of complex tissues.

\paragraph{Dataset forms on performance} A key observation is the varying model performance across different dataset forms (D1, D2, and D3). The D1 form of the dataset (full image) is relatively limited in size, with only 147 images. On the other hand, the D2 dataset (patching) significantly expands the sample size to 76,413 patches, while D3 (superpixel) consists of 4,039 superpixels. The substantial difference in dataset sizes suggests that models trained on D2 likely benefit from the abundance of training data, especially for tissues like granulation, necrosis, and slough. This could explain why classification models performed better on D2 than on D3. The D2 dataset likely had a better balance or clearer representation of tissue classes, whereas D3 might suffer from class imbalance or less informative features, leading to reduced performance. In D1, bone and tendon tissues are poorly represented (only 12 and 15 samples, respectively), and even in D2 and D3, these classes are heavily underrepresented (D2: 240 bones, 448 tendons; D3: 18 bones, 80 tendons). This class imbalance likely plays a significant role in the poor performance of segmentation and classification models on bone and tendon tissues across all dataset forms. The models do not have enough training examples to effectively learn from these rare classes.
This suggests that the dataset's characteristics play a critical role in model performance, emphasising the need for careful dataset curation and balancing when training classification models, especially in medical imaging tasks.

\paragraph{Granulation and necrotic tissue success} The consistent performance for granulation and necrotic tissues suggests that these models are well-tuned for tissue types with distinct textural patterns, making them easier to differentiate. This strength may be attributed to the combination of dataset features and the architecture of the models, which are designed to capture such patterns effectively. However, overfitting to these classes might explain the lack of generalisation to other tissue types like bone and tendon.

\paragraph{Failure in bone and tendon recognition} The poor performance in bone and tendon classification across models, especially for D3, highlights future work opportunities. Bone and tendon tissues may share similar intensities with other tissues, such as slough, or may not have enough distinct features for the models to detect reliably. The failure in these classes could be due to the lack of sufficient training examples or misrepresentation of these tissues in the training set. Future improvements could focus on augmenting the dataset or applying advanced techniques like domain adaptation to enhance bone and tendon recognition.

\paragraph{Segmentation vs. classification approach} The main strength of segmentation models is their ability to generate pixel-wise predictions, which is crucial for medical imaging tasks like wound tissue analysis, where spatial information and tissue boundaries are paramount. The segmentation models' higher IoU and Dice scores for granulation and necrotic tissues highlight their ability to handle intricate wound textures and details. These models utilize spatial hierarchies and deep feature extraction techniques to capture and localize tissues more effectively. The results show that segmentation models are better suited for medical tasks requiring tissue delineation, which is essential in wound assessment and treatment planning.

On the other hand, classification models are typically faster and simpler to implement but lack the fine-grained spatial understanding necessary for segmenting complex tissue types. While they performed well in assigning tissue labels to the overall image, their inability to focus on precise regions within the wound area limits their utility for detailed medical diagnosis. They showed strong performance for certain tissues, such as granulation and necrosis, but struggled with more nuanced or less-represented classes like bone and tendon. 

\subsection{Limitations}
The primary limitation of this study stems from the small and imbalanced nature of the dataset. The under-representation of certain tissue types, particularly bone and tendon, significantly impacted the performance of both segmentation and classification models, leading to poor generalisation for these tissue classes. This imbalance poses a challenge for the models to learn sufficient features for accurate tissue recognition and segmentation, resulting in low-performance metrics for these underrepresented classes. Additionally, the limited size of the dataset restricts the models' ability to generalise to new, unseen data, potentially affecting the reliability of the results when applied in broader clinical settings.

Another limitation arises from the heterogeneity in image quality. The wound images were collected from various sources, each with different levels of resolution, lighting conditions, and image clarity. This variability in image quality introduces noise and inconsistency, which complicates the training process and may hinder the model's ability to learn robust features. The lack of standardised image acquisition protocols across the dataset may lead to suboptimal performance in real-world applications where high consistency in imaging is required for reliable wound assessment.

Also, while meticulous effort was made to label wound tissues accurately, agreement among expert physicians on wound labelling can vary substantially, making it challenging to establish a consistent criterion standard~\cite{jamanetworkopen2021}. This variability could impact the interpretability and reproducibility of results, particularly when deploying models in diverse clinical environments.

Despite these challenges, this study marks a significant first step toward leveraging publicly available wound images to develop a labelled wound tissue segmentation dataset. It highlights the feasibility of using these resources to create a labelled dataset for model development, paving the way for future efforts to enhance the dataset in size and quality. 

\subsection{Future directions}
\paragraph{Augmentation, synthetic data generation and data balancing}
The poor performance of the models on underrepresented tissue classes, particularly bone and tendon, highlights the need for enhanced data augmentation, synthetic data generation, and balancing techniques. Advanced augmentation methods such as rotation, flipping, zooming, and colour adjustments can increase the diversity of training samples, exposing models to more varied scenarios and improving their robustness in imbalanced datasets~\citep{shorten2019survey}. Also, synthetic data generation can be a powerful technique to augment rare classes like bone and tendon by creating artificial but realistic images using methods such as the generative adversarial networks (GANs) technique. These synthetic images can help balance the dataset and provide the models with more examples to learn from, especially for underrepresented tissue types~\citep{zhu2017unpaired}.
Alongside these techniques, oversampling or SMOTE (Synthetic Minority Over-sampling Technique) can further address class imbalance, ensuring that the model trains on a more evenly distributed dataset and reducing its bias toward overrepresented classes~\citep{chawla2002smote}. By combining traditional augmentation, synthetic data generation, and data balancing methods, the dataset's representation of bone and tendon can be significantly improved, resulting in better segmentation performance for these challenging tissues.

\paragraph{Custom loss functions}
A promising approach to further improve the recognition of underrepresented tissues is the use of custom loss functions. By assigning higher weights to under-performing classes like bone and tendon, models can be more heavily penalised for misclassifications, encouraging them to focus on these harder-to-classify tissues. Loss functions such as focal loss are particularly effective for addressing class imbalance by down-weighting easy examples and emphasising difficult ones~\citep{Lin2017ICCV}. This method has significantly improved datasets with skewed class distributions~\citep{zhang2018generalized}.

\paragraph{Ensemble and custom models}
The varied success of different model architectures across tissue classes suggests that ensemble modelling could be a viable solution to mitigate these discrepancies. By combining the strengths of multiple models—such as UNet++ 2D, FPN+VGG16, and LinkNet+VGG16—an ensemble approach can take advantage of each model's strengths while compensating for their weaknesses. Techniques such as stacking, bagging, or voting allow multiple models to contribute to the final prediction, improving accuracy and robustness, particularly in complex tasks like wound segmentation~\citep{dietterich2000ensemble}. Ensemble models have been widely used in medical image analysis to achieve state-of-the-art results and can be especially effective for improving performance on poorly represented tissue types~\citep{ganaie2022ensemble}.

One potential approach for a custom model would be to incorporate multi-scale feature extraction modules, allowing the model to capture high-level contextual information and fine details, essential for accurate wound tissue delineation~\citep{poudel2021deep}. Also, attention mechanisms could be integrated to help the model focus on important tissue regions, potentially improving its accuracy on challenging classes~\citep{sinha2020multi,su2021msu}. Recent studies (e.g.,~\citep{niu2021hybrid,guo2022segnext}) have shown that attention-based modules can enhance segmentation performance by guiding the model's focus to regions with higher relevance, which would be particularly useful in medical image analysis where certain tissue classes need more emphasis.

\section{Conclusion}\label{sec:conclusion}
This study makes significant contributions by presenting the first publicly available wound tissue segmentation dataset, offering a valuable resource for the research community focused on wound analysis and medical imaging. A comprehensive comparison of deep learning (DL) models for wound tissue segmentation and classification was conducted, exploring their performance across different dataset forms, including full image, patch, and superpixel representations. By developing and testing 82 DL models, this study provides an in-depth evaluation of model performance across six critical tissue types: slough, granulation, maceration, necrosis, bone, and tendon.

One of the major contributions of this work is its rigorous evaluation of model performance using multiple metrics, including precision, recall, Dice score, IoU, AUC and MCC. These comprehensive evaluations allow for a nuanced understanding of each model's strengths and limitations. The results revealed that modified segmentation models, particularly FPN+VGG16, consistently outperformed conventional models, showing their superiority over the full image dataset in handling complex tissue boundaries and structures.

The significance of this study lies not only in its immediate findings but also in its potential as a benchmark for future research. It also provides extensive discussion highlighting the gaps and potential future work, particularly in addressing the segmentation of bone and tendon tissues. The publicly available dataset and extensive model evaluations can serve as a foundation for further developments in wound tissue segmentation and classification. Researchers can use this dataset to test new algorithms, refine existing models, and explore innovative techniques to improve accuracy, particularly for underrepresented tissue types like bone and tendon. This work sets a strong precedent for addressing real-world challenges in medical imaging, providing both practical insights and a valuable resource for continued advancements in the field.

\section*{Data availability statement}
This study utilised publicly available secondary data. The labelled dataset created in this study is partially available at \url{https://github.com/akabircs/WoundTissue}. The full labelled dataset will be available after acceptance.
% The data supporting the findings of this study are available in the GitHub repository at the following URL: https://github.com/MattPears1/AI-Expert-Urology-DoubleBlind-Study2024.

\section*{Ethics statement}
This study utilised publicly available secondary data, and we adhered to relevant terms and conditions associated with its use. No new data were collected directly from human participants. Our study complies with all ethical guidelines and regulations concerning the responsible use of secondary data.

\section*{Declaration of competing interest}
The authors declare that they have no conflict of interest.

\section*{Funding}
This research did not receive any specific grant from funding agencies in the public, commercial, or not-for-profit sectors.

% \section*{CRediT authorship contribution statement}
% \textbf{Ashad Kabir:} Conceptualization, Methodology, Formal analysis, Software, Data Curation, Writing - Original Draft,  Writing - Review \& Editing, Visualization, Validation, Project administration. \textbf{Nidita Roy:} Methodology, Formal analysis, Software, Data Curation. \textbf{Md. Ekramul Hossain:} Formal analysis, Writing - Original Draft, Visualization. \textbf{Jill Featherston:} Data Curation, Validation, Writing - Review \& Editing. \textbf{Sayed Ahmed:} Conceptualization, Writing - Review \& Editing.

% \bibliographystyle{elsarticle-harv}
\bibliographystyle{elsarticle-num-names}
\bibliography{reference}

\end{document}